\documentclass[amsmath,amssymb,prl,aps,superscriptaddress,
floatfix,tightenlines,showpacs,twocolumn
]{revtex4-2}
\usepackage[T1]{fontenc}
\usepackage[latin9]{inputenc}
\usepackage{babel}
\usepackage{calc}
\usepackage{amsmath}
\usepackage{amssymb}
\usepackage{braket}
\usepackage{graphicx}
\usepackage{subfigure}
\usepackage{color}
\usepackage[unicode=true]
 {hyperref}

\def\ket#1{|#1\rangle}
\def\avg#1{\langle#1\rangle}

\def\be{\begin{equation}}       \def\ee{\end{equation}}
\def\bea{\begin{eqnarray}}      \def\eea{\end{eqnarray}}
\def\bp{\begin{pmatrix}} \def\ep{\end{pmatrix}}
\def\beaa{\begin{equation}\begin{aligned}}
\def\eeaa{\end{aligned}\end{equation}}

\def\nn{\nonumber}

\makeatletter

\providecommand{\LyX}{\texorpdfstring
  {L\kern-.1667em\lower.25em\hbox{Y}\kern-.125emX\@}
  {LyX}}

\makeatother

\begin{document}
\preprint{APS/123-QED}
\title{Kinetic Energy Driven Ferromagnetic Insulator}
\author{Jinyuan Ye}
\affiliation{ Department of Physics, Fudan University, Shanghai, 200433, China}
\affiliation{New Cornerstone Science Laboratory, Department of Physics, School of Science, Westlake University, Hangzhou 310024, Zhejiang, China}
\affiliation{Institute of Natural Sciences, Westlake Institute for Advanced Study, Hangzhou 310024, Zhejiang, China}
\author{Yuchi He}
\email{yuchi.he@physics.ox.ac.uk}
\affiliation{Rudolf Peierls Centre for Theoretical Physics, Clarendon Laboratory, Parks Road, Oxford OX1 3PU, United Kingdom}
\author{Congjun Wu}
\email{wucongjun@westlake.edu.cn}
\affiliation{New Cornerstone Science Laboratory, Department of Physics, School of Science, Westlake University, Hangzhou 310024, Zhejiang, China}
\affiliation{Institute of Natural Sciences, Westlake Institute for Advanced Study, Hangzhou 310024, Zhejiang, China}
\affiliation{Institute for Theoretical Sciences, Westlake University, Hangzhou 310024, Zhejiang, China}
\affiliation{Key Laboratory for Quantum Materials of Zhejiang Province, School of Science, Westlake University, Hangzhou 310024, Zhejiang, China}

\begin{abstract}
We construct a minimal model of interacting fermions establishing a ferromagnetic insulating phase.
It is based on the Hubbard model on a trimerized triangular lattice in the regime of $t\gg |t^\prime|>0$ with $t$ and $t^\prime$ the intra- and inter-trimer hopping amplitudes, respectively.
At the $\frac{1}{3}$-filling, each trimer becomes a triplet spin-1 moment, and the inter-trimer superexchange is ferromagnetic with  
$J =- \frac{2}{27}\frac{t^{\prime 2}}{t}$ in the limit of $U/t=+\infty$. 
As $U/t$ becomes finite, the antiferromagnetic superexchange competes with the ferromagnetic one.
The system enters into a frustrated antiferromagnetic insulator when $\lambda >U/t\gg 1$ where $\lambda$ is a constant
at the order of 10.
In contrast, a similar analysis to the trimerized Kagome lattice shows that only the antiferromagnetic superexchange exists
at $\frac{1}{3}$-filling.
\end{abstract}
\maketitle

The mechanism of ferromagnetism (FM) is a long-standing problem of strong correlation physics~\cite{kanamori1963}. 
The driving force of
itinerant FM is often thought to be the direct exchange among electrons with the same spin to reduce the inter-particle repulsion. 
Spin polarization pays a large cost of kinetic energy due to Pauli's exclusion principle, such that in most situations, fermions would rather develop 
unpolarized but highly correlated many-body ground states than be polarized.
Nevertheless, a few rigorous results have been established:
The Nagaoka theorem proves that the infinite $U$ Hubbard model at half-filling with doping only a single hole develops FM \cite{nagaoka1966,haerter2005}, for a bipartite lattice regardless of the sign of hopping, and for a non-bipartite lattice with positive hopping.
Another class of theorems of FM rely on the flat-band structure of line graphs in which the kinetic energy is suppressed to zero\cite{mielke1992,mielke1993,mielke1993a,tasaki1995}.
It is also shown that FM could remain stable under certain conditions even when the band structure becomes non-singular\cite{tasaki1996}.
Furthermore, a series of theorems are proved that Hund's interaction combined with the quasi-1D band structure lead to itinerant FM in the multi-orbital Hubbard model over a large region of filling factors \cite{liwu2014}.
The Curie-Weiss metal state and the FM criticality are accurately studied by quantum Monte-Carlo simulations free of the sign problem \cite{xuwu2015}.

On the other hand, Mott insulators are typically dominated by antiferromagnetic (AFM) superexchange.
Upon doping, they may serve as the parent compounds of high $T_c$ superconductors. 
In frustrated systems, such as triangular and Kagome lattices, the AFM spin alignment of each bond can not be simultaneously satisfied due to the geometry constraint. 
An incredibly rich and complex nature of quantum magnetism manifests
~\cite{mott1949,ANDERSON1973,lee2006,norman2016,kanoda2011mott}, leading to exotic states of spin liquid
~\cite{Sachdev2004,vasiliev_milestones_2018,lucile_2017}.
Owing to the complexity of frustrated magnets, the cluster model approach is widely employed for theoretical studies. 
It extends beyond the concept of individual sites, focusing instead on well-defined clusters of atoms as the fundamental units. 
Unequal coupling strengths cause electrons to localize 
inside these clusters, rather than on individual atomic sites. 
The localized degrees of freedom can be effectively described in terms of molecular orbitals, giving rise to what are often referred to as \textit{molecules in solids}~\cite{khomskii_orbital_2021}.
Coupled clusters can form cluster Mott insulators~\cite{yao2010,he2018,hu_li_2023,jiang2020,giri2017}.
Experimentally, such clusters in triangular/Kagome lattices (trimers) have been observed in materials such as $\text{Nb}_3\text{Cl}_8$~\cite{grytsiuk2024}, $\text{LiVO}_2$~\cite{pen1997}, and the $\text{Mo}_3\text{O}_8$ family of compounds~\cite{nikolaev_quantum_2021,Akbari2018,chen2018,gall_synthesis_2013,torardi_synthesis_1985}. 

It would be non-trivial to unify FM and AFM  in the same system regarding to their very different origins. 
In this article, we find the transition from an AFM insulator to an
FM one simply as increasing interaction strengths. 
It is based on the Hubbard model defined in a trimerized triangular lattice at the $\frac{1}{3}$-filling in the strong coupling regime $U\gg t \gg |t^\prime|$.
Two electrons in each trimer form a spin-1 triplet at $t>0$.
The inter-trimer hopping generates both FM and AFM superexchanges: The former involves intermediate excitations free of double occupancy with the energy cost of $t$ analogous to the case of charge-transfer insulators, 
while the latter generates double occupancy with the energy cost of $U$ similar to the case of Mott insulators.
At $U/t=+\infty$, the FM exchange dominates while it switches to the AFM one at around $U/t\approx 13\sim 15$, as confirmed by our density-matrix-renormalization-group~\cite{White1992,tenpy, PhysRevLett.99.127004,pollmann2009,zauner2015transfer, rams2018} (DMRG) simulations. 
The FM insulating state remains robust by threading a weak staggered flux pattern of $\phi$, and it becomes a FM metal at small 
doping levels. 
In contrast, as for the trimerized Kagome lattice, the inter-trimer exchange is always AFM-like. 
Our mechanism is different from the orbital-active Mott insulators in which FM exchange could appear according to  
the Kugel-Khomskii physics \cite{kugel1982}.
In that case, the overall nature of superexchange remains anti-ferro, either ferro in orbital and antiferro in spin, {\it or}, antiferro in orbital and ferro in spin.

\begin{figure}[t]
\centering
\includegraphics[width=0.8\columnwidth]{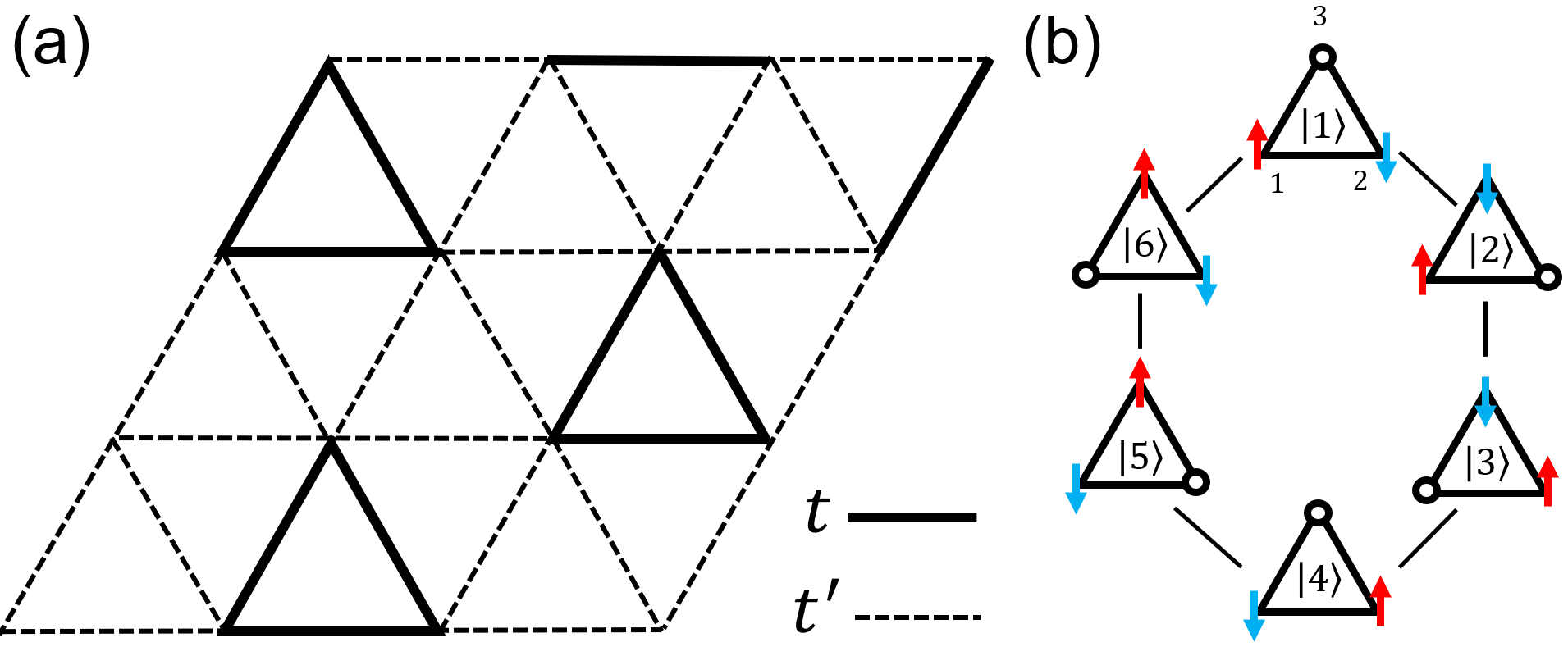}
\caption{(a) The trimerized
triangular lattice with hopping strengths \textemdash 
the intra-trimer hopping $t$ and the inter-trimer hopping $t^\prime$ represented by the solid and dashed lines, respectively. 
Each trimer is filled with 2 electrons. 
(b) The bases of the sector with $S_{\text{tot}}=0$ are generated by the hole's hopping around the trimer, sequentially  denoted as 
$\ket{1}=c_{1\uparrow}^\dagger c_{2\downarrow}^\dagger \ket{\Omega},
\ket{2}=c_{3\downarrow}^\dagger c_{1\uparrow}^\dagger \ket{\Omega}, 
\ket{3}=c_{2\uparrow}^\dagger c_{3\downarrow}^\dagger \ket{\Omega},
\ket{4}=c_{1\downarrow}^\dagger c_{2\uparrow}^\dagger \ket{\Omega},
\ket{5}=c_{3\uparrow}^\dagger c_{1\downarrow}^\dagger \ket{\Omega}$, and 
$\ket{6}=c_{2\downarrow}^\dagger c_{3\uparrow}^\dagger \ket{\Omega}$.
}
\label{Fig:lattice}
\end{figure}

\textit{Model Hamiltonian.}\textemdash
We consider the Hubbard model $H=H_0+H^\prime$ defined in a trimerized triangular lattice as illustrated in Fig.~\ref{Fig:lattice} (a),
\bea
H_0&=&t\sum_{\avg{ij}} \left\{ c^\dagger_{i\sigma}c_{j\sigma}+h.c. \right \} 
+ \sum_i U n_{i,\uparrow}n_{i,\downarrow}, \nn \\
H^\prime&=& t^\prime \sum_{\avg{\avg{i^\prime j^\prime}}}
\left\{ c^\dagger_{i^\prime \sigma}c_{j^\prime \sigma}+h.c. \right \},
\label{eq:Ham}
\eea 
where $H_0$ is the intra-trimer Hamiltonian, and $H^\prime$ describes the inter-trimer hopping; $\avg{ij}$ and $\avg{\avg{i^\prime j^\prime}}$ represent the intra- and inter- trimer bonds, respectively.
To establish FM below half-filling, it is crucial to have $t>0$ due to the absence of particle-hole symmetry in non-bipartite lattices. 
Two neighboring trimers are connected by two links, hence, the superexchanges between them 
are insensitive to the sign of $t^\prime$.

The free band structure of the trimerized configuration is depicted in Supplemental Material (SM) Sect. A~\cite{supp}.
Since $t\gg |t^\prime|$, three intra-trimer states are solved as orbitals:
The lower two are degenerate with the energy of $E=-t$ and the upper with $E=2t$.
They are broadened into three bands whose widths are proportional to $|t^\prime|$.
The lower two bands overlap and are separated from the upper one. 
The band gap $\Delta_b$ between them is at the order of $t$ at $|t^\prime/t|\ll 1$, while it closes at $t^\prime/t=3/4$.

\begin{figure}[t]
\centering
\includegraphics[width=0.7\linewidth]{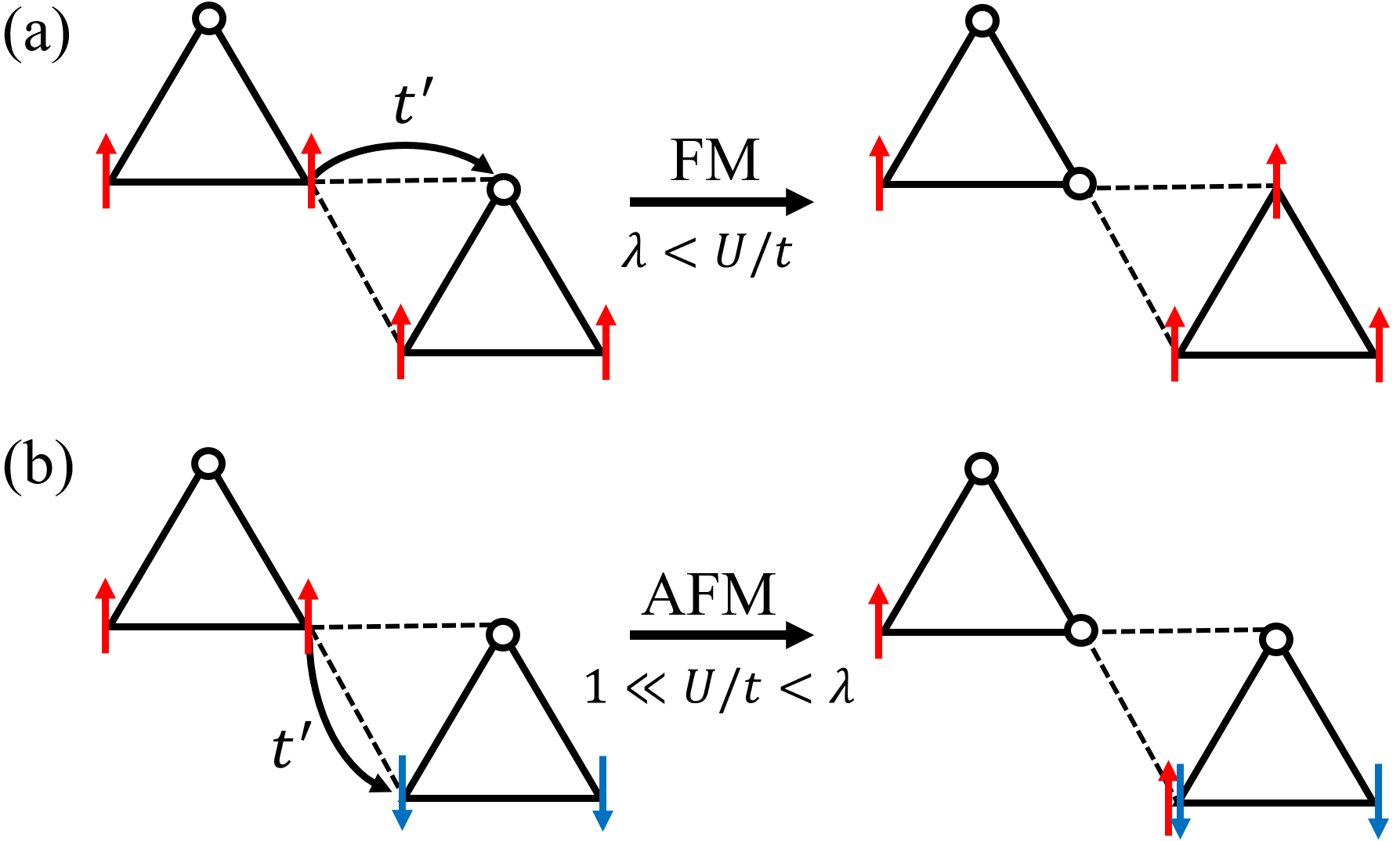}
\caption{Particle-hole excitations leading to (a) the FM  superexchange at $U/t >\lambda$ and (b) the AFM one at $\lambda > U/t \gg 1$, respectively, with the transition value 
$\lambda\sim 10$.
For the FM case, the excitation energy $E_{ex}\sim t$ with only single occupations
analogues to charge-transfer insulators,
while $E_{ex}\sim U$ for the AFM case manifesting the Mott physics characterized by the double occupancy. 
}
\label{fig:FM vs AFM}
\end{figure}

Below we consider the commensurate filling of $\frac{1}{3}$, {\it i.e.}, 2 fermions per trimer.
The lower two bands are effectively half-filled, hence, it should be metallic in the weak coupling regime of $U/t^\prime  \lesssim 1$. 
In contrast, the strong coupling regime is characterized by $\min(U, t)\gg t^\prime$, in which the correlated insulating states appear. 
The inter-trimer hopping generates virtual excitations involving 1 or 3 fermions in a trimer. 
The dependence of $E_{ex}$ on $t$ and $U$ here are depicted in Figs.~\ref{fig:FM vs AFM}(a)~and~\ref{fig:FM vs AFM}(b), respectively (see the SM Sect. B~\cite{supp}).
Excitations with $E_{ex}\sim t$ lead to the FM superexchange in the regime of 
$U\gg t\gg t^\prime$.
As lowering $U/t$, the AFM superexchange becomes dominant as in the usual Mott insulators, and the excitations are characterized by double occupancy with $E_{ex}\sim U$.
The evolution from the FM to AFM superexchange is explained below.

We begin with the case of $U/t = +\infty$ which forbids the double occupancy. 
The Nagaoka theorem applies to the case with only a single hole in the entire system \cite{nagaoka1966}.
Below the dominance of the FM exchange is shown to occur at the $\frac{1}{3}$-filling, however.

\textit{Single trimer.}\textemdash 
At $t^\prime=0$ the system is reduced to disconnected trimers. 
When the trimer is filled with 2 electrons,
it is sufficient to consider the sector of $S_{\text{tot},z}=0$ within a trimer based on the SU(2) symmetry.
This local Hilbert space contains 6 bases denoted as $\ket{m} ~(m=1\sim 6)$: $\ket{1}=c_{1\uparrow}^\dagger c_{2\downarrow}^\dagger \ket{\Omega}$ with $\ket{\Omega}$ denoting the vacuum state, and the other states $\ket{m}$ are generated as the hole hops around the trimer in a clockwise way as shown in Fig.~\ref{Fig:lattice} (b).
In this convention,  $\ket{m}$ and $\ket{m+3}$ correspond to a pair of states by flipping two spins.
Applying $H_0$ on $|m\rangle$, it yields 
$H_0\ket{m}=-t(\ket{m-1}+\ket{m+1})$, where $m$ is defined modulo 6.

Remarkably, this intra-trimer 2-electron problem exhibits a 6-fold rotational symmetry.
This can be mapped to a rolling motion problem: 
The pair of spin represents a 2-tooth external gear, and the trimer behaves as a 3-tooth internal gear.
Since rolling is a combination of translation and rotation, one round is insufficient to restore the configurations of both degrees of freedom back to the initial ones. 
Instead, the minimal requirement to a periodicity requires rolling two rounds. 
Consequently, this problem is mapped to a single-body problem moving around a hexagon. 
It is interesting that the effective orbital angular momentum here is modulo 6 instead of 3, exhibiting a fractionalization behavior. 

The eigenstates for the single trimer problem in the sector of  $S_{\text{tot},z}=0$
are solved as 
\begin{equation}
\ket{k_n}=\frac{1}{\sqrt{6}}\sum_n e^{ik_n m}\ket{m},
\label{eq:intra-trimer}
\end{equation}
with the energy spectrum, 
$    E_n=-2t\cos{k_n}$,
where $k_n=\frac{n}{3} \pi$ with $n=0,\pm 1,\pm 2,3$.
The states with $n=0, \pm 2$ are symmetric under the operation $m\to m+3$, hence, they are spin-triplet. 
In contrast, the other states with odd values of $n$ are spin-singlet. 
For $t>0$, the ground state with $n=0$ is a spin triplet, and the lowest excitations are a pair of singlets with a gap of $t$.

\textit{Trimerized triangular lattice.} \textemdash  
Next consider the case of $0<|t'|/t\ll 1$.
Since the trimer ground state is spin-1, $H^\prime$ generates inter-trimer superexchanges and lifts the degeneracy. 
The weakly-coupled trimers are effectively described by the spin-1 Heisenberg model in a triangular lattice where one site represents a trimer. 
The effective exchange Hamiltonian reads,
\begin{equation}
H_{ex}=J\sum_{\langle ij \rangle}
\mathbf{S}_i \cdot \mathbf{S}_j +C ,
\label{eq:FMexchange}
\end{equation}
where $\mathbf{S}_i$ represents the total
spin of the trimer $i$;
$J$ is the exchange energy and $C$
is an energy constant, which will be determined below. 

The $J$ turns out to be FM as calculated via the 2nd order degenerate perturbation theory. 
The total spin of two neighboring trimers lie in 3 channels of $S_{\mathrm{tot}}=2,1$, and $0$.
The energy gains in these channels are shown in SM Sect. C~\cite{supp}, yielding that
$\Delta E^{(2)}/(\frac{t^{\prime 2}}{t})= -\frac{10}{27},-\frac{6}{27}, -\frac{4}{27}$, respectively. 
Comparing to Eq.~\eqref{eq:FMexchange}, we arrive at
\bea
J=-\frac{2}{27} \frac{{t^\prime}^2}{t}, \ \ 
C=4J.
\label{eq:J}
\eea
Therefore, the $\frac{1}{3}$-filled trimerized triangular lattice is in the FM insulating state in the limit of $U\to \infty$ and $t^\prime/t\ll 1$ due to the FM superexchange.
 

\textit{Flux threading.}\textemdash
Next consider the effect of a trimerized triangular lattice with a staggered flux pattern $\pm\phi$ threading each plaquette as shown in SM. Sect. C~\cite{supp}.
Correspondingly, the intra-trimer hopping amplitude $t$ and the inter-trimer one $t^\prime$ are modified as $t e^{\pm i \frac{\phi}{3} }$ and $t^\prime e^{\pm i\frac{\phi}{3}}$, where the signs $\pm$ are determined by whether the hopping is along or against the flux winding, respectively.
The flux modifies the energy spectrum of the 6 intra-trimer two-electron states as
\begin{equation}
    E_n=-2t\cos\left (k_n+\frac{\phi}{3}\right ),
    \label{eq:flux}
\end{equation}
where $\phi$ is defined modulo $2\pi$.
Nevertheless, the spectral flow indicated by Eq.~\eqref{eq:flux} shows that the dispersion returns back at $\phi =\pi$, \textit{i.e.}, $n\to n+1$ (\mbox{mod} 6), but switching triplet and singlet states.


The FM exchange described in Eq.~\eqref{eq:FMexchange} remains robust at small values of $|\phi|< \frac{\pi}{2}$ since the intra-trimer ground state remains the triplet $|k_0\rangle$ .
Nevertheless, $J$ is reduced:
The 2nd order perturbation calculation shows that at $|\phi|< \frac{\pi}{2}$ the flux dependence of $J(\phi)$ reads
\bea
J(\phi)\approx J\left(1-\frac{7}{54}\phi^2\right).
\label{eq:Jtheta}
\eea
A more accurate expression is obtained as
$J(\phi)=-\frac{{t^\prime}^2}{18t}
 \left( \cos\frac{2\phi}{3} \cos\frac{\phi}{3}\right )/
\left( \cos(\frac{\phi}{3}+\frac{\pi}{6})  \cos(\frac{\phi}{3}-\frac{\pi}{6}) \right)$.

When $\phi$ reaches $\pm \frac{\pi}{2}$, the intra-trimer singlet and triplet states become degenerate.
According to the decomposition rule of the SU(2) representations, two trimers result in 6 spin channels: 
one set of quintet ($S_{\mathrm{tot}}=2$), three sets of triplet  ($S_{\mathrm{tot}}=1$), and two sets of singlet ($S_{\mathrm{tot}}=0$). 
The lowest and highest energy states are both spin singlets (see the SM Sect. C~\cite{supp}): The former is the direct-product state of two trimer singlets, and the latter is the entangled one built up by two trimer triplets, exhibiting the energies of 
$\Delta E^{(2)}_{s,1}= -\frac{83{t^\prime}^2}{108\sqrt{3}t}$ and
$\Delta E^{(2)}_{s,2}= -\frac{5 {t^\prime}^2}{12\sqrt{3}t}$, respectively.
They do not mix at the level of the 2nd order perturbation theory, nevertheless, a small mixing could occur at a high order.
This means that as increasing $|\phi|$ from $0$ to $\frac{\pi}{2}$ the ground state  of the entire lattice changes from the spin fully polarized state to the direct product state of the singlet of each trimer. 
The transition between two different types of ground states will be deferred to a future research. 

\textit{Phase transitions by tuning $t'$ and $U$.}\textemdash
We now consider finite values of $U$, which generates the competition between AFM and FM exchanges. 
For simplicity only the case of $\phi=0$
is considered here. 
The intra-trimer ground states are a set of spin-1 triplet as long as $U>0$~\cite{supp},
in contrast to the case of a square that the ground state becomes spin-$\frac{3}{2}$ at a large value of $U/t\ge 18.7$~\cite{dehollain2020nagaoka}.
The detailed calculation of the 
superexchange coupling $J$ between two neighboring spin-1 trimers at finite values of $U$ is presented in SM. Sect. D~\cite{supp}, 
yielding
\bea
J=-\frac{2}{27} \frac{t^{\prime 2}}{t} + 
\frac{62}{81} \frac{t^{\prime 2}}{U}.
\label{eq:fullexchange}
\eea
where $1\ll U/t<\infty$ is assumed.
The 1st term in Eq. (\ref{eq:fullexchange}) arises from the FM super-exchange as explained before, and the 2nd one is the AFM super-exchange involving doubly occupied intermediate states. 
The overall value of $J$ switches from FM to AFM approximately at 
$U_F/t\approx 10.3$ at which $J=0$ 
Around $J=0$, high-order superexchanges would be important.


\begin{figure}[t]
\centering
\includegraphics[width=0.9\linewidth]{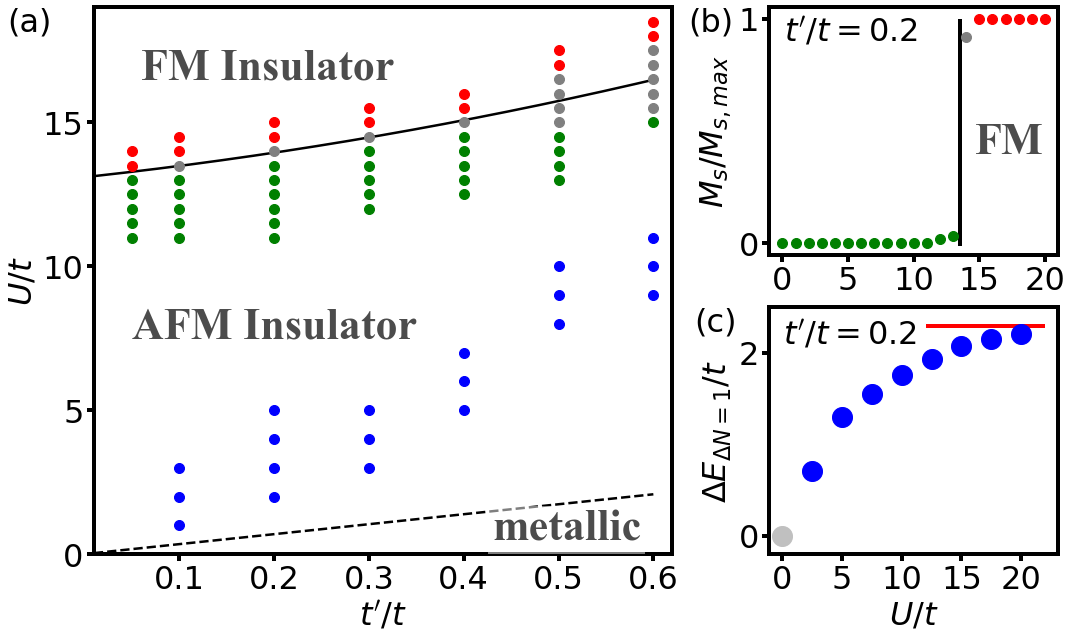}
\caption{
(a) The phase diagram at the $\frac{1}{3}$-filling for $0 < t^\prime/t \le 0.6$. 
DMRG calculations are applied for the system on a tilted cylinder with the size of $L_x \times L_y \times 3$ to determine the boundary between the fully spin-polarized FM insulating state and the unpolarized AFM one, which is marked by a solid line. 
The AFM-FM transition is determined based on $L_x=6, L_y=4$. 
The data of $L_x=\infty$ and $L_y=3$ are also calculated yielding similar results. 
The decay length $\xi_e$ is extracted from the single-electron Green's function by extending the system to the quasi-1D limit of $L_x \to \infty$ (see SM. Sect. G).
Charge localizations are found down to certain values of $U$, and the data with $\xi_e$ less than the inter-trimer distance $\sqrt 3 a$,  are marked by blue dots for reference. 
The metal-insulator transition boundary (black dashed line) is estimated from the relation $U_c / W \approx \sqrt{3}/2$, where $W$ is the free bandwidth. 
(b) The relative spontaneous magnetization $M_s / M_{\text{s},\max}$, with $M_s = \sqrt{\langle \boldsymbol{S}_{\text{tot}}^2 \rangle}$ and $M_{\text{s},\max} = \sqrt{S_{\max} (S_{\max} + 1)}$ where $S_{\max} = L_x\times L_y$ equals to the number of trimers.
It reaches 1 at $U/t \approx 14$, confirming full polarization.
(c) The single-electron gap, defined as $\Delta = E_{N=49}^{\text{GS}} + E_{N=47}^{\text{GS}} - 2E_{N=48}^{\text{GS}}$, exhibits non-zero values except at $U=0$. The $U=0$ result has exact zero gap because there are degenerate single-particle states at Fermi energy.
The red solid line is the band gap when electrons are fully spin polarized for comparison. 
The deviation in FM region is due to the fact that spins remain fully polarized for doping a hole but partially flip when adding an electron. 
}
\label{Fig:DMRG}
\end{figure}


We compute the phase diagram of the trimerized triangular lattice model at the $\frac{1}{3}$-filling in $t^\prime/t$-$U/t$ parameter space, as shown in Fig.~\ref{Fig:DMRG} (a).
At $t^\prime/t\ge3/4$, the band gap closes and the system remains metallic even at large values of $U/t$. 
Our main purpose here is to study the competition between FM and AFM superexchanges in the insulating states and leave the metallic state for future studies, hence, only the parameter range of $0<t^\prime/t\le 0.6$  will be explored here.
DMRG simulations are performed to a 72-site system. The FM superexchange dominates over the AFM one, leading to a fully polarized FM insulating state when $U/t$ is larger than a $t^\prime/t$-dependent critical value $U_F/t$ as indicated by the solid line. 
As $t^\prime$ increases, higher order virtual 
hopping processes become more prominent and then larger values of $U_F/t$ are required to establish the FM phase. 
As an example, the transition to spin polarized states at $t^\prime/t=0.2$ is shown in Fig.~\ref{Fig:DMRG} (b), indicating $U_F/t\approx 14$, and the corresponding single-electron gap $\Delta$ is shown in Fig.~\ref{Fig:DMRG} (c). 
There may exist intermediate phases near the transition, which will be deferred to future studies. 

As further lowering $U/t$, the system is expected to transition into the gapless metallic phase in the thermodynamic limit. 
The precise location of this phase boundary is challenging due to the intrinsic difficulty of many-body simulations. 
For this we employ the iDMRG method for a quasi-1D system. 
The equal-imaginary-time single-electron Green's functions decay exponentially in the AFM state, whose decaying length $\xi_e$'s are extracted as shown in SM Sect. I~\cite{supp}.
As further lowering the interaction strength below the critical value of $U_c/t$, the single-electron gap closes and the system enters the metallic phase. 
According to the self-consistent solution to the equation of motion of Green's function~\cite{hubbard1963electron}, an estimation of $U_c/W\gtrsim\sqrt{3}/2$ is expected as marked in Fig.~\ref{Fig:DMRG}(a), where $W$ is the band width determined by $t^\prime/t$ (see SM. Sect. E).
Certainly, this method only yields a rough estimation which cannot fully capture the correlation effect. 
A precise determination of the phase boundary of the metal-insulator transition at $\frac{1}{3}$-filling is beyond the scope of this work.


\textit{Discussions.}\textemdash
We explore the universality underlying
the above mechanism to FM,
including the formation of the molecular-type (intra-trimer) high-spin moment, and the
inter-molecular (inter-trimer) FM coherence arising from the kinetic energy gain via superexchange interactions. 
When two fermions fill the doubly degenerate intra-trimer levels, they form a spin-1 moment at $U>0$. 
This Hund's rule type physics is a reminiscence of the flat-band FM in the limit of few sites.
Trimer molecules become Mott-insulating when the inter-trimer hopping $|t^\prime|\ll t, U$. 
The kinetic energy is lowered due to the inter-trimer virtual hoppings leading to superexchange interactions. 
If the intermediate excitations generated by virtual hoppings are dominated by doubly occupied sites, the resulting superexchange is AFM-like. 
Since our molecular Mott-insulators are at $\frac{1}{3}$-filling, when $U\gg t$ the dominant excitations are actually free of double occupancy with excitation energies at the order of $t$, which is similar to charge-transfer insulators, leading to the FM superexchange.

The FM nature of superexchanges in our case becomes manifest by comparing the AFM and FM configurations of two neighboring spin-1 trimers: $S_z=\pm 1$ for each trimer, respectively in the former case, and $S_z=1$ for both trimers in the latter. 
In both cases, the intermediate excitations are free of double occupations, consisting of a 3-filled trimer and a singly filled one.
In the AFM configuration, the virtual hopping processes lead to different spin configurations.
They are incoherent and thus unable to optimize the kinetic energy. 
In contrast, similar to the Nagaoka FM, all virtual hopping processes in the case of FM configuration lead to the same fully polarized intermediate state, hence, all virtual hopping processes are coherent such that the kinetic energy gain is maximized.  
Hence, our description constitutes a hybrid, generalized mechanism that extends beyond the strict requirement of either Nagaoka or flat-band FM models.
Experimentally, in multi-orbital solid state systems, Hund's coupling often exists promoting the formation of local high-spin moments.
When there exist other low energy orbitals to host spin polarized intermediate excitations generated by virtual hopping processes, our FM mechanism can also apply to such kind of materials.

It would be interesting to further explore the physics by doping the $\frac{1}{3}$-filling correlated insulating state in the triangular lattice. 
When the FM superexchange dominates, the system will become a FM metal upon slight hole doping.
The doped holes move in the background of spin-1 moments coupled by the FM super-exchange, which still results in FM polarization. 
At $\frac{1}{4}$-filling, {\it i.e.}, the case that the average fermion number in each trimer equals $\frac{3}{2}$, our preliminary simulation results show that for the value of $t^\prime/t=0.2$, the system has already been fully spin polarized as a FM metal at $U_F/t>10$, which is notably lower than that at $\frac{1}{3}$-filled insulating state.
The $\frac{1}{4}$-filling can be viewed as half of the spin-1 trimers are replaced by spin-1/2 fermions, and the itineracy facilitates the FM coherence~\cite{future_study}.
The detailed study will be deferred for a future publication. 

We briefly discuss the situation of $t<0$. 
For a single trimer, the 2-particle ground state 
is a spin-singlet corresponding to $n=3$ in Eq.~\eqref{eq:intra-trimer}.
There exists a gap to the intra-trimer triplet excitations at $\Delta=t$.
Hence, the effective model at $\frac{1}{3}$-filling is a rotor model with the inter-trimer coupling at the order of $t^\prime$ yielding non-magnetic ground state in the regime of $t^\prime\ll t$. 
On the other hand, our previous studies apply to 
the case of the $\frac{2}{3}$-filling, since it is mapped to the $\frac{1}{3}$-filling at $t>0$ by a particle-hole transformation.

\textit{Conclusion.}\textemdash 
We propose a mechanism to the FM insulating state in the trimerized triangular lattice, which occurs at the $\frac{1}{3}$-filling with $t>0$ in the regime of $U\gg t\gg |t^\prime|$.
In each trimer, two electrons form spin-1 moments due to the ``orbital'' degeneracy and repulsive interaction. 
The inter-trimer hopping generates superexchange couplings to lower the kinetic energy.
At $U/t\to \infty$, only the FM superexchange exists, which is weakened by introducing a staggered flux pattern.
As $U/t$ becomes finite, both the FM and AFM superexchanges contribute, and the former wins over the latter around $U/t\gtrsim 15$.
FM metallic state may appear upon slightly doping the $\frac{1}{3}$-filled FM insulating state. 
This work provides valuable insights into the study of quantum magnetism in strongly correlated fermion systems. 

\vspace{4mm}
We are grateful to Z. M. Pan, Y. Wang, K. Yang and J. Zhang for valuable discussions.
CW is supported by the National Natural Science Foundation of China under the Grant Nos. 12574274, 12234016, and 12550402.
YH is supported by European Research Council (ERC) under the European Union Horizon 2020 Research and Innovation Programme (Grant Agreement Nos. 804213-TMCS). The QuSpin~\cite{phillip2019} and TeNPy~\cite{tenpy} packages were used for the numerical studies. This work has been supported by the New Cornerstone Science Foundation. The computation resource was provided by the Westlake HPC Center.

\bibliographystyle{prsty}
\bibliography{ref}

\section*{End Matter}

\begin{figure}[h]
    \centering
\includegraphics[width=0.9\linewidth]{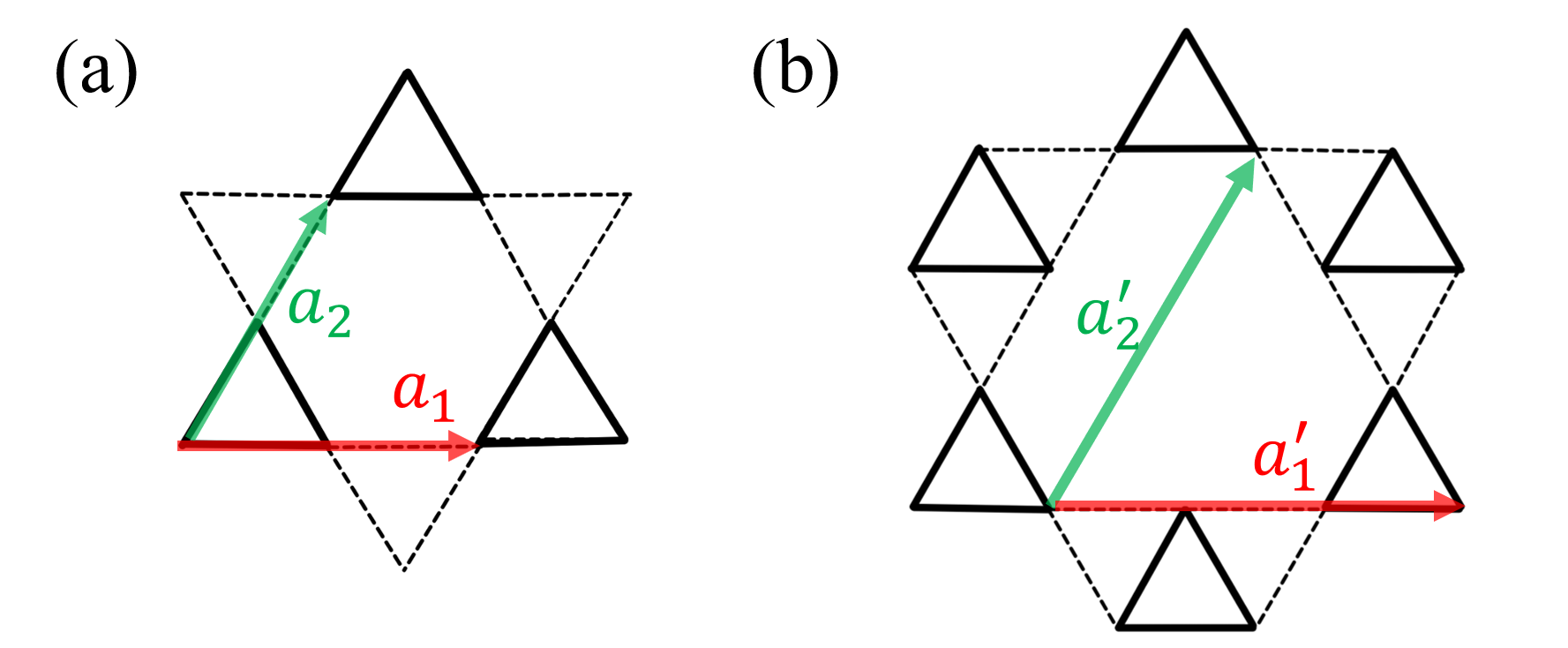}
\caption{~(a) The trimerized Kagome lattice in which trimers form a triangular lattice.  
 (b) The Kagome trimer lattice. 
 Two neighboring trimers are connected via one bond  in (a) and two bonds in (b), respectively.}
\label{fig:kagome}
\end{figure}

\textit{Extension.}\textemdash 
The studies in the main text can be easily generalized to other trimerized lattices and for simplicity, only the case of $t^\prime\ll t<U$ is discussed below.
Fig.~\ref{fig:kagome} (a) shows the trimerized Kagome lattice (breathing Kagome lattice) at the $\frac{1}{3}$-filling, in which two neighboring trimers only connect via one inter-trimer hopping.
In the case of $t>0$, the lowest band is flat with a quadratic touching with the 2nd one.
The widths of the 2nd and 3rd bands are at the order of $|t^\prime|$, and these bands are separated by the gap $\Delta_b\sim t$.
When the flat-band is half-filled, \textit{i.e.}, the $\frac{1}{6}$-filling, the system is FM which should be stable with respect to small doping away from the $\frac{1}{6}$-filling.
Nevertheless, as shown in SM Sect. D~\cite{supp}, the super-exchange between two neighboring trimers at the $\frac{1}{3}$-filling in the regime $U\gg t\gg t^\prime$ is always AFM. 
In fact, the exchange disappears at $U/t=\infty$ because no spin-flip process takes place via a single link connecting two trimers. 
Finite $U$ renders spin-flip processes feasible yielding the same Hamiltonian as Eq.~\eqref{eq:J} with the AFM exchange $J=\frac{44}{81} \frac{t^{\prime 2}}{U}$.
Since the trimers form a triangular lattice, the system will exhibit the 120$^\circ$ pattern of the AFM order.

The lattice shown in Fig.~\ref{fig:kagome}(b) is dubbed 
the Kagome trimer lattice, where trimers form a Kagome lattice and the connection between neighboring trimers is via two bonds, hence, the superexchange will be the same as 
in Eq.~\eqref{eq:fullexchange}.
Therefore, as increasing $U/t$, the effective model will also undergo a transition from the spin-1 AFM Heisenberg model to the FM one in the Kagome lattice. 

\textit{Possible intermediate states}\textemdash
The super-exchange $J$ in Eq. (\ref{eq:fullexchange})
between two neighboring spin-1 trimers changes sign in the insulating states, leading to the transition from 
the AFM to FM states as increasing $U/t$. 
Around $J=0$, high-order exchanges including the bi-quadratic exchange, or, ring exchange terms, could play an important role and lead to novel quantum magnetic intermediate states.
These states in the thermodynamic limit may exhibit novel AFM orderings, partial FM polarization, or even more exotic quantum paramagnetic states, {\it etc}. 
Our DMRG data indicate that there could exist a  partially polarized phase, but we cannot exclude the possibility whether the observation is a finite-size effect constrained by computational resources. 
It would be very interesting to further explore the possible intermediate phases, which is highly non-trivial. Nevertheless, the main point of this manuscript is to provide a robust mechanism to establish a kinetic energy driven FM-state. 
A comprehensive study of the intermediate magnetic states is a matter of accurately simulating a 2D strongly correlated fermion model, which is  beyond the scope of this work, which will be deferred to future studies.

As for the metal-AFM insulator transition at a lower value of $U/t$, it is a notoriously difficult problem of 2D strongly correlated systems in the absence of particle-hole symmetry. 
For the case of $t^\prime=t$, the metal-insulator transition in the triangular lattice remains an open issue under debate in the research community~\cite{Tocchio2021, szasz2020}. 
Our trimerized case actually enriches research in this direction since the local moments in the trimerized clusters are of spin-1 instead of spin-1/2. 
Again, a comprehensive study of this issue is out of the scope of the current manuscript, and will be deferred for future studies.

\end{document}


\preprint{APS/123-QED}
\title{Kinetic Energy Driven Ferromagnetic Insulator}

\author{Jinyuan Ye}
\affiliation{ Department of Physics, Fudan University, Shanghai, 200433, China}
\affiliation{New Cornerstone Science Laboratory, Department of Physics, School of Science, Westlake University, Hangzhou 310024, Zhejiang, China}
\affiliation{Institute of Natural Sciences, Westlake Institute for Advanced Study, Hangzhou 310024, Zhejiang, China}
\author{Yuchi He}
\email{yuchi.he@physics.ox.ac.uk}
\affiliation{Rudolf Peierls Centre for Theoretical Physics, Clarendon Laboratory, Parks Road, Oxford OX1 3PU, United Kingdom}
\author{Congjun Wu}
\email{wucongjun@westlake.edu.cn}
\affiliation{New Cornerstone Science Laboratory, Department of Physics, School of Science, Westlake University, Hangzhou 310024, Zhejiang, China}
\affiliation{Institute of Natural Sciences, Westlake Institute for Advanced Study, Hangzhou 310024, Zhejiang, China}
\affiliation{Institute for Theoretical Sciences, Westlake University, Hangzhou 310024, Zhejiang, China}
\affiliation{Key Laboratory for Quantum Materials of Zhejiang Province, School of Science, Westlake University, Hangzhou 310024, Zhejiang, China}

\maketitle
\tableofcontents
\renewcommand{\theequation}{S\arabic{equation}}
\setcounter{equation}{0}
\renewcommand{\thefigure}{S\arabic{figure}}
\setcounter{figure}{0}
\renewcommand{\thetable}{S\arabic{table}}
\setcounter{table}{0}

\section{Non-interacting band spectrum of the trimerized triangular lattice}
\label{App:numericalsolution}

\begin{figure}[t]
    \centering
    \subfigure[$t^\prime/t=0.2$]{\includegraphics[width=0.15\columnwidth]{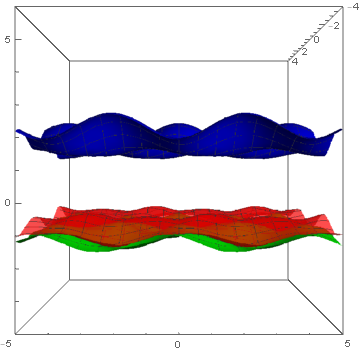}}
    \subfigure[$t^\prime/t=0.6$]{\includegraphics[width=0.15\columnwidth]{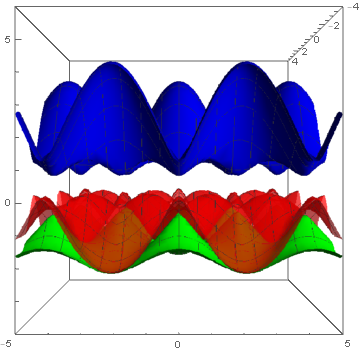}}
    \subfigure[$t^\prime/t=0.8$]{\includegraphics[width=0.15\columnwidth]{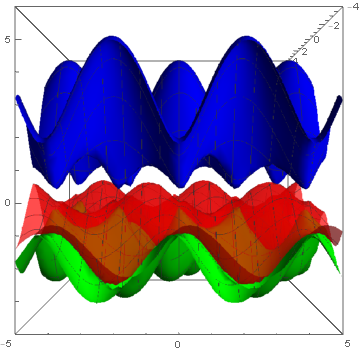}}
    \subfigure[$t^\prime/t=1$]{\includegraphics[width=0.15\columnwidth]{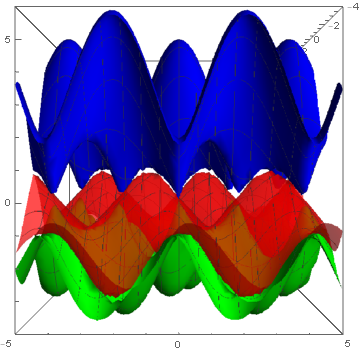}}\\
    
    \caption{The non-interacting band structure for various $t'/t$}
    \label{Fig:spectrum}
\end{figure}

The calculation details of the non-interacting band structure are shown below.
The non-interacting Hamiltonian in the Bloch basis is:
\begin{equation}
    \begin{pmatrix}
        0
        &t+t^\prime \left(e^{-i\vec{k}\cdot\vec{a}_1}+e^{-i\vec{k}\cdot\vec{a}_2}\right)
        &t+t^\prime e^{-i\vec{k}\cdot\vec{a}_1}\left(1+e^{i\vec{k}\cdot\vec{a}_2}\right)\\
        t+t^\prime \left(e^{i\vec{k}\cdot\vec{a}_1}+e^{i\vec{k}\cdot\vec{a}_2}\right)
        &0
        &t+t^\prime e^{i\vec{k}\cdot\vec{a}_2}\left(1+e^{-i\vec{k}\cdot\vec{a}_1}\right)\\
        t+t^\prime e^{i\vec{k}\cdot\vec{a}_1}\left(1+e^{-i\vec{k}\cdot\vec{a}_2}\right)
        &t+t^\prime e^{-i\vec{k}\cdot\vec{a}_2}\left(1+e^{i\vec{k}\cdot\vec{a}_1}\right)
        &0\\
    \end{pmatrix}
\end{equation}
where $\vec{k}=(k_x,k_y)$ and $\vec{a}_1=(0,\sqrt{3}),\vec{a}_2=(\frac{3}{2},\frac{\sqrt{3}}{2})$. 
For disconnected trimers ($t^\prime=0$), the matrix reduces to the form:
\begin{equation}
    \begin{pmatrix}
        0&t&t\\
        t&0&t\\
        t&t&0
    \end{pmatrix}
\end{equation} 
and the spectrum consists of three flat bands: $\varepsilon_1 = \varepsilon_2 = -t$, $\varepsilon_3 = 2t$. 
For finite $t^\prime$, the eigenvalues satisfy a cubic equation
$$\varepsilon^3-p\varepsilon+q=0$$
where 
$$
\begin{aligned}
    p&=\left\lvert t+t^\prime \left(e^{-i\vec{k}\cdot\vec{a}_1}+e^{-i\vec{k}\cdot\vec{a}_2}\right)\right\rvert^2+2\left\lvert t+t^\prime e^{-i\vec{k}\cdot\vec{a}_1}\left(1+e^{i\vec{k}\cdot\vec{a}_2}\right)\right\rvert^2
    \\
    q&=-2\left\lvert t+t^\prime e^{-i\vec{k}\cdot\vec{a}_1}\left(1+e^{i\vec{k}\cdot\vec{a}_2}\right)\right\rvert^2 \text{Re}\left(t+t^\prime \left(e^{-i\vec{k}\cdot\vec{a}_1}+e^{-i\vec{k}\cdot\vec{a}_2}\right)\right)
\end{aligned}$$
Applying the cubic formula
$$
\begin{aligned}
    \varepsilon_1&=\left(\frac{q}{2}+\sqrt{\frac{q^2}{4}-\frac{p^3}{27}}\right)^{\frac{1}{3}}+\left(-\frac{q}{2}+\sqrt{\frac{q^2}{4}-\frac{p^3}{27}}\right)^{\frac{1}{3}}\\
    \varepsilon_2&=e^{i\frac{2\pi}{3}}\left(\frac{q}{2}+\sqrt{\frac{q^2}{4}-\frac{p^3}{27}}\right)^{\frac{1}{3}}+e^{-i\frac{2\pi}{3}}\left(-\frac{q}{2}+\sqrt{\frac{q^2}{4}-\frac{p^3}{27}}\right)^{\frac{1}{3}}\\
    \varepsilon_3&=e^{-i\frac{2\pi}{3}}\left(\frac{q}{2}+\sqrt{\frac{q^2}{4}-\frac{p^3}{27}}\right)^{\frac{1}{3}}+e^{i\frac{2\pi}{3}}\left(-\frac{q}{2}+\sqrt{\frac{q^2}{4}-\frac{p^3}{27}}\right)^{\frac{1}{3}}\\
\end{aligned}
$$
the non-interacting band dispersion is calculated for various $t^\prime/t$ ratios, as depicted in Fig.~\ref{Fig:spectrum}.
The bandwidth is proportional to $t^\prime$, and the gap closes at $t^\prime/t=3/4$.

\section{Single trimer at finite $U$}
\label{App:solution of single trimer}

As a starting point, the Hamiltonian $H_0$ has been diagonalized for a single trimer filled with 1 electron (1/6-filling), 2 electrons (1/3-filling), and 3 electrons(half-filling).

\subsection{$\frac{1}{6}$-filling}
\label{App:eigenstate of trimer at n=1}

When the trimer is filled with 1 electron, it reduces to a single-electron problem, where the on-site interaction does not contribute.
Due to the SU(2) symmetry of Hamiltonian $H_0$, it is enough to consider the $S_z=\frac{1}{2}$ sector.
The trimer has $C_3$ symmetry, and this symmetry allows us to classify the $S_z=\frac{1}{2}$ states using the basis
$\ket{q_{-1}=-\frac{2\pi}{3}}=\frac{e^{-i2\pi/3}c_{1\uparrow}^\dagger+e^{i 2\pi/3}c_{2\uparrow}^\dagger+c_{3\uparrow}^\dagger}{\sqrt{3}}\ket{\Omega},
\ket{q_{1}=\frac{2\pi}{3}}=\frac{e^{i2\pi/3}c_{1\uparrow}^\dagger+e^{-i2\pi/3}c_{2\uparrow}^\dagger+c_{3\uparrow}^\dagger}{\sqrt{3}}\ket{\Omega},
\ket{q_0=0}=\frac{c_{1\uparrow}^\dagger+c_{2\uparrow}^\dagger+c_{3\uparrow}^\dagger}{\sqrt{3}}\ket{\Omega}$
which are exactly the eigenstates.
The corresponding eigenenergies are $-t,-t,2t$.
The eigenstates in the $S_z=-\frac{1}{2}$ sector are degenerate with those in the $S_z=\frac{1}{2}$ sector.
Hence, we will not distinguish the two spin sectors (e.g.$S_z=\frac{1}{2}\ \text{and}\ S_z=-\frac{1}{2}$) below, if there is no ambiguity.

The solution in the presence of flux threading is similar. 
Flux does not break the $C_3$ symmetry; therefore, the eigenstates are the same but with shifted eigenenergies:
$2t\cos{\left(\theta-\frac{2}{3}\pi\right)},2t\cos{\left(\theta+\frac{2}{3}\pi\right)},2t\cos{\theta}$,
where we define $\phi/3=\theta$ for simplicity. 
This convention is maintained throughout this Supplementary Material.

\subsection{$\frac{1}{3}$-filling}
\label{App:eigenstate of trimer at n=2}

When one trimer is filled with 2 electrons, the situation is quite different, as on-site interaction matters in the Hamiltonian.

The spin SU(2) symmetry of the Hamiltonian allows us to focus on the $S_z=0$ sector.  The Hilbert space dimension of the $S_z=0$ sector is 9. 
Then, according to $C_3$ symmetry, the Hilbert space can be divided into three sectors:
\begin{equation}
\begin{pmatrix}
    0&-2t&2t\\
    -2t&0&-2t\\
    2t&-2t&U\\
\end{pmatrix},
\begin{pmatrix}
    0&-t&te^{i\frac{\pi}{3}}\\
    -t&0&te^{-i\frac{\pi}{3}}\\
    te^{-i\frac{\pi}{3}}&te^{i\frac{\pi}{3}}&U\\
\end{pmatrix},
\begin{pmatrix}
    0&-t&te^{-i\frac{\pi}{3}}\\
    -t&0&te^{i\frac{\pi}{3}}\\
    te^{i\frac{\pi}{3}}&te^{-i\frac{\pi}{3}}&U\\
\end{pmatrix}
\end{equation}
with the following basis states which are separated by a good quantum number $k$:
\begin{equation}
\begin{aligned}
    &\ket{\varphi_1,k=0}=\frac{1}{\sqrt{3}}\left(c_{1\uparrow}^\dagger c_{2\downarrow}^\dagger+c_{2\uparrow}^\dagger c_{3\downarrow}^\dagger+c_{3\uparrow}^\dagger c_{1\downarrow}^\dagger\right)\ket{\Omega}\\
    &\ket{\varphi_2,k=0}=\frac{1}{\sqrt{3}}\left(c_{1\downarrow}^\dagger c_{2\uparrow}^\dagger+c_{2\downarrow}^\dagger c_{3\uparrow}^\dagger+c_{3\downarrow}^\dagger c_{1\uparrow}^\dagger\right)\ket{\Omega}\\
    &\ket{\varphi_3,k=0}=\frac{1}{\sqrt{3}}\left(c_{1\uparrow}^\dagger c_{1\downarrow}^\dagger+c_{2\uparrow}^\dagger c_{2\downarrow}^\dagger+c_{3\uparrow}^\dagger c_{3\downarrow}^\dagger\right)\ket{\Omega}\\
    &\ket{\varphi_1,k=\frac{2\pi}{3}}=\frac{1}{\sqrt{3}}\left(c_{1\uparrow}^\dagger c_{2\downarrow}^\dagger+e^{i\frac{2\pi}{3}}c_{2\uparrow}^\dagger c_{3\downarrow}^\dagger+e^{-i\frac{2\pi}{3}}c_{3\uparrow}^\dagger c_{1\downarrow}^\dagger\right)\ket{\Omega}\\
    &\ket{\varphi_2,k=\frac{2\pi}{3}}=\frac{1}{\sqrt{3}}\left(c_{1\downarrow}^\dagger c_{2\uparrow}^\dagger+e^{i\frac{2\pi}{3}}c_{2\downarrow}^\dagger c_{3\uparrow}^\dagger+e^{-i\frac{2\pi}{3}}c_{3\downarrow}^\dagger c_{1\uparrow}^\dagger\right)\ket{\Omega}\\
    &\ket{\varphi_3,k=\frac{2\pi}{3}}=\frac{1}{\sqrt{3}}\left(c_{1\uparrow}^\dagger c_{1\downarrow}^\dagger+e^{i\frac{2\pi}{3}}c_{2\uparrow}^\dagger c_{2\downarrow}^\dagger+e^{-i\frac{2\pi}{3}}c_{3\uparrow}^\dagger c_{3\downarrow}^\dagger\right)\ket{\Omega}\\
    &\ket{\varphi_1,k=-\frac{2\pi}{3}}=\frac{1}{\sqrt{3}}\left(c_{1\uparrow}^\dagger c_{2\downarrow}^\dagger+e^{-i\frac{2\pi}{3}}c_{2\uparrow}^\dagger c_{3\downarrow}^\dagger+e^{i\frac{2\pi}{3}}c_{3\uparrow}^\dagger c_{1\downarrow}^\dagger\right)\ket{\Omega}\\
    &\ket{\varphi_2,k=-\frac{2\pi}{3}}=\frac{1}{\sqrt{3}}\left(c_{1\downarrow}^\dagger c_{2\uparrow}^\dagger+e^{-i\frac{2\pi}{3}}c_{2\downarrow}^\dagger c_{3\uparrow}^\dagger+e^{i\frac{2\pi}{3}}c_{3\downarrow}^\dagger c_{1\uparrow}^\dagger\right)\ket{\Omega}\\
    &\ket{\varphi_3,k=-\frac{2\pi}{3}}=\frac{1}{\sqrt{3}}\left(c_{1\uparrow}^\dagger c_{1\downarrow}^\dagger+e^{-i\frac{2\pi}{3}}c_{2\uparrow}^\dagger c_{2\downarrow}^\dagger+e^{i\frac{2\pi}{3}}c_{3\uparrow}^\dagger c_{3\downarrow}^\dagger\right)\ket{\Omega}\\
\end{aligned}
\nonumber
\end{equation}
The Hamiltonian of the trimer with 2 electrons has almost been diagonalized.
Note that as long as $U>0,t>0$, the ground state is a triplet.

In the following discussion, among those eigenstates belonging to one trimer filled with two electrons, only the ground state is required. 
For the $S_z=0$ sector, it reads: 
\begin{equation}
\begin{aligned}
    \ket{k_0,0}&=\frac{1}{\sqrt{2}}\left(\ket{\varphi_1,k=0}+\ket{\varphi_2,k=0}\right)\\
    &=\frac{1}{\sqrt{6}}\left(c_{1\uparrow}^\dagger c_{2\downarrow}^\dagger+c_{1\downarrow}^\dagger c_{2\uparrow}^\dagger+c_{2\uparrow}^\dagger c_{3\downarrow}^\dagger+c_{2\downarrow}^\dagger c_{3\uparrow}^\dagger+c_{3\uparrow}^\dagger c_{1\downarrow}^\dagger+c_{3\downarrow}^\dagger c_{1\uparrow}^\dagger\right)\ket{\Omega}\text{, }\\
\end{aligned}
\end{equation}
which is a triplet, equivalent to a spin-1 moment, and the other two states are
\begin{equation}
\begin{aligned}
    &\ket{k_0,1}=S_{+}\ket{k_0,0}=\frac{1}{\sqrt{3}}\left(c_{1\uparrow}^\dagger c_{2\uparrow}^\dagger+c_{2\uparrow}^\dagger c_{3\uparrow}^\dagger+c_{3\uparrow}^\dagger c_{1\uparrow}^\dagger\right)\ket{\Omega}\\
    &\ket{k_0,-1}=S_{-}\ket{k_0,0}=\frac{1}{\sqrt{3}}\left(c_{1\downarrow}^\dagger c_{2\downarrow}^\dagger+c_{2\downarrow}^\dagger c_{3\downarrow}^\dagger+c_{3\downarrow}^\dagger c_{1\downarrow}^\dagger\right)\ket{\Omega}\\
\end{aligned}
\end{equation}
with an eigenenergy $-2t$.

\subsection{Half-filling}
\label{App:eigenstate of trimer at n=3}
Similarly, the SU(2) symmetry allows us to focus on  $S_z=\frac{1}{2}$.  Using $C_3$ symmetry, the Hamiltonian is reduced to the direct sum of three $3\times3$ matrices, which is written below:
\begin{equation}
    \begin{pmatrix}
        0&&\\
        &U&\\
        &&U\\
    \end{pmatrix},\ \begin{pmatrix}
        0&\sqrt{3}t e^{-i\frac{5\pi}{6}}&\sqrt{3}t e^{i\frac{\pi}{2}}\\
        \sqrt{3}t e^{i\frac{5\pi}{6}}&U&\sqrt{3}t e^{-i\frac{\pi}{6}}\\
        \sqrt{3}t e^{-i\frac{\pi}{2}}&\sqrt{3}t e^{i\frac{\pi}{6}}&U\\
    \end{pmatrix},\ \begin{pmatrix}
        0&\sqrt{3}t e^{i\frac{5\pi}{6}}&\sqrt{3}t e^{-i\frac{\pi}{2}}\\
        \sqrt{3}t e^{-i\frac{5\pi}{6}}&U&\sqrt{3}t e^{i\frac{\pi}{6}}\\
        \sqrt{3}t e^{i\frac{\pi}{2}}&\sqrt{3}t e^{-i\frac{\pi}{6}}&U\\
    \end{pmatrix},
\end{equation}
and the basis states are listed below, with the good quantum number $S_z=\frac{1}{2}$ and $k=0,\pm\frac{2\pi}{3}$, which reads:
\begin{equation}
\begin{aligned}
    &\ket{\psi_1,k=0}=\frac{1}{\sqrt{3}}\left(c_{1\uparrow}^\dagger c_{2\uparrow}^\dagger c_{3\downarrow}^\dagger+c_{1\uparrow}^\dagger c_{2\downarrow}^\dagger c_{3\uparrow}^\dagger+c_{1\downarrow}^\dagger c_{2\uparrow}^\dagger c_{3\uparrow}^\dagger\right)\ket{\Omega}\\
    &\ket{\psi_2,k=0}=\frac{1}{\sqrt{3}}\left(c_{1\uparrow}^\dagger c_{1\downarrow}^\dagger c_{2\uparrow}^\dagger+c_{2\uparrow}^\dagger c_{2\downarrow}^\dagger c_{3\uparrow}^\dagger+c_{3\uparrow}^\dagger c_{3\downarrow}^\dagger c_{1\uparrow}^\dagger\right)\ket{\Omega}\\
    &\ket{\psi_3,k=0}=\frac{1}{\sqrt{3}}\left(c_{1\uparrow}^\dagger c_{1\downarrow}^\dagger c_{3\uparrow}^\dagger+c_{2\uparrow}^\dagger c_{2\downarrow}^\dagger c_{1\uparrow}^\dagger+c_{3\uparrow}^\dagger c_{3\downarrow}^\dagger c_{2\uparrow}^\dagger\right)\ket{\Omega}\\
    &\ket{\psi_1,k=\frac{2\pi}{3}}=\frac{1}{\sqrt{3}}\left(c_{1\uparrow}^\dagger c_{2\uparrow}^\dagger c_{3\downarrow}^\dagger+e^{i\frac{2\pi}{3}}c_{1\uparrow}^\dagger c_{2\downarrow}^\dagger c_{3\uparrow}^\dagger+e^{-i\frac{2\pi}{3}}c_{1\downarrow}^\dagger c_{2\uparrow}^\dagger c_{3\uparrow}^\dagger\right)\ket{\Omega}\\
    &\ket{\psi_2,k=\frac{2\pi}{3}}=\frac{1}{\sqrt{3}}\left(c_{1\uparrow}^\dagger c_{1\downarrow}^\dagger c_{2\uparrow}^\dagger+e^{i\frac{2\pi}{3}}c_{2\uparrow}^\dagger c_{2\downarrow}^\dagger c_{3\uparrow}^\dagger+e^{-i\frac{2\pi}{3}}c_{3\uparrow}^\dagger c_{3\downarrow}^\dagger c_{1\uparrow}^\dagger\right)\ket{\Omega}\\
    &\ket{\psi_3,k=\frac{2\pi}{3}}=\frac{1}{\sqrt{3}}\left(c_{1\uparrow}^\dagger c_{1\downarrow}^\dagger c_{3\uparrow}^\dagger+e^{i\frac{2\pi}{3}}c_{2\uparrow}^\dagger c_{2\downarrow}^\dagger c_{1\uparrow}^\dagger+e^{-i\frac{2\pi}{3}}c_{3\uparrow}^\dagger c_{3\downarrow}^\dagger c_{2\uparrow}^\dagger\right)\ket{\Omega}\\
    &\ket{\psi_1,k=-\frac{2\pi}{3}}=\frac{1}{\sqrt{3}}\left(c_{1\uparrow}^\dagger c_{2\uparrow}^\dagger c_{3\downarrow}^\dagger+e^{-i\frac{2\pi}{3}}c_{1\uparrow}^\dagger c_{2\downarrow}^\dagger c_{3\uparrow}^\dagger+e^{i\frac{2\pi}{3}}c_{1\downarrow}^\dagger c_{2\uparrow}^\dagger c_{3\uparrow}^\dagger\right)\ket{\Omega}\\
    &\ket{\psi_2,k=-\frac{2\pi}{3}}=\frac{1}{\sqrt{3}}\left(c_{1\uparrow}^\dagger c_{1\downarrow}^\dagger c_{2\uparrow}^\dagger+e^{-i\frac{2\pi}{3}}c_{2\uparrow}^\dagger c_{2\downarrow}^\dagger c_{3\uparrow}^\dagger+e^{i\frac{2\pi}{3}}c_{3\uparrow}^\dagger c_{3\downarrow}^\dagger c_{1\uparrow}^\dagger\right)\ket{\Omega}\\
    &\ket{\psi_3,k=-\frac{2\pi}{3}}=\frac{1}{\sqrt{3}}\left(c_{1\uparrow}^\dagger c_{1\downarrow}^\dagger c_{3\uparrow}^\dagger+e^{-i\frac{2\pi}{3}}c_{2\uparrow}^\dagger c_{2\downarrow}^\dagger c_{1\uparrow}^\dagger+e^{i\frac{2\pi}{3}}c_{3\uparrow}^\dagger c_{3\downarrow}^\dagger c_{2\uparrow}^\dagger\right)\ket{\Omega}\\
\end{aligned}
\nonumber
\end{equation}
Note that the states belonging to the $k=0$ sectors do not mix up with each other.

For our purposes, the Hamiltonian is diagonalized in the limit of $U\gg t$ as follows:
\begin{equation}
\begin{aligned}
    &\ket{\psi_1,0}&0\\
    &\ket{\psi_1,\pm\frac{2\pi}{3}}&-\frac{6t^2}{U}\\
    &\ket{\psi_2,0}&U\\
    &\ket{\psi_3,0}&U\\
    &\ket{\psi_{\pm},\frac{2\pi}{3}}=\frac{1}{\sqrt{2}}\left(\ket{\psi_2,\frac{2\pi}{3}}\pm e^{i\frac{\pi}{6}}\ket{\psi_3,\frac{2\pi}{3}}\right)&U\pm\sqrt{3}t\\
    &\ket{\psi_{\pm},-\frac{2\pi}{3}}=\frac{1}{\sqrt{2}}\left(\ket{\psi_2,-\frac{2\pi}{3}}\pm e^{-i\frac{\pi}{6}}\ket{\psi_3,-\frac{2\pi}{3}}\right)&U\pm\sqrt{3}t\\
\end{aligned}
\label{eq:eigenstate for n=3}
\end{equation}
where eigenstate is written on the left, and corresponding eigenenergies on the right.

Within the solution, $\ket{\psi_1,0},\ket{\psi_2,0},\ket{\psi_3,0}$ and the corresponding energy are exact solutions regardless of $U$.
For $\ket{\psi_1,\pm\frac{2\pi}{3}}$, we neglect the 1st-order corrections to the wavefunction and the energy is calculated by 2nd-order perturbation theory, under the condition $U\gg t$.
Furthermore, the energies of $\ket{\psi_\pm,\pm\frac{2\pi}{3}}$ can be approximated as $U$ under the condition $U\gg t$.
This approximation enables us to treat $c_{i\uparrow}^\dagger c_{i\downarrow}^\dagger c_{j\uparrow}^\dagger\ket{\Omega},i\neq j$ as eigenstates directly, significantly simplifying all the following analytical calculations.
The remaining eigenstates in other spin sectors can be obtained via SU(2) symmetry.

In this section, we solve the Hamiltonian for a single trimer below and at half-filling, with the convention of positive $t$. In the next section, we will introduce the details of 2nd-order perturbation theory based on the solution of the single trimer.

\section{The 2nd order degenerate perturbation theory}

In this section, we present a systematic analysis of effective exchange interaction between two trimers (A \& B) under both staggered flux and flux-free conditions in the $U\to\infty$ limit. 
The unperturbed ground states $\ket{\psi_0}$ -- formed as direct products of $\ket{k_0,0(\pm1)}_A$ and $\ket{k_0,0(\pm1)}_B$.
This results in 9 degenerate configurations, all with an unperturbed eigenenergies of $E^{(0)}_0=-4t$. 
Again, the SU(2) symmetry inherent in the Hamiltonian allows the decomposition of these unperturbed ground states into three distinct sectors, which are characterized by total spin quantum numbers $S = 2, 1, 0$.

Within each sector, the energy corrections induced by the perturbation $H'$ remain identical. 
We therefore select a representative state from each sector for computational efficiency:
\begin{align*}
\ket{S=2} &= \ket{k_0,1}_A \otimes \ket{k_0,1}_B, \\
\ket{S=1} &= \frac{1}{\sqrt{2}}\left(\ket{k_0,1}_A \otimes \ket{k_0,0}_B - \ket{k_0,0}_A \otimes \ket{k_0,1}_B\right), \\
\ket{S=0} &= \frac{1}{\sqrt{3}}\left(\ket{k_0,1}_A \otimes \ket{k_0,-1}_B - \ket{k_0,0}_A \otimes \ket{k_0,0}_B + \ket{k_0,-1}_A \otimes \ket{k_0,1}_B\right).
\end{align*}
These state labeled with different $S$ allows us to employ 2nd-order perturbation theory to calculate the second-order energy corrections:
\begin{equation}
    \Delta E^{(2)} = \sum_{m} \frac{|\bra{m} H' \ket{\psi_0}|^2}{E^{(0)}_0 - E^{(0)}_m},
    \label{eq:2nd_perturbation}
\end{equation}
where $\ket{\psi_0}$ denotes the representative states $\ket{S=2,1,0}$, and $\ket{m}$ runs over the complete set of normalized intermediate states generated through $H'\ket{\psi_0}$.

\subsection{Flux-free case}
\label{App:energy without flux}

We first consider the case where no flux threads through the lattice system.
For the $S=2$ channel, the unperturbed state $\ket{S=2}$ possesses an eigenenergy of $-4t$. Introducing inter-trimer hopping induces an additional energy lowering, quantified through 2nd-order perturbation theory (Eq.~\eqref{eq:2nd_perturbation}).

The intermediate states for the $S=2$ channel are generated through:
\begin{equation}
    H^\prime\ket{S=2}=-\frac{t^\prime}{3}\left(\sqrt{2}c_{1^\prime\uparrow}^\dagger c_{2^\prime\uparrow}^\dagger c_{3^\prime\uparrow}^\dagger\frac{c_{2\uparrow}^\dagger-c_{1\uparrow}^\dagger}{\sqrt{2}}+2\sqrt{2}c_{1\uparrow}^\dagger c_{2\uparrow}^\dagger c_{3\uparrow}^\dagger\frac{c_{2^\prime\uparrow}^\dagger-c_{1^\prime\uparrow}^\dagger}{\sqrt{2}}\right)\ket{\Omega}
\end{equation}
These intermediate configurations consist of trimer pairs with (1,3) electron occupations, exemplified by: $\ket{m_1} = c_{1'\uparrow}^\dagger c_{2'\uparrow}^\dagger c_{3'\uparrow}^\dagger \frac{c_{2\uparrow}^\dagger - c_{1\uparrow}^\dagger}{\sqrt{2}}\ket{\Omega}, 
\ket{m_2} = c_{1\uparrow}^\dagger c_{2\uparrow}^\dagger c_{3\uparrow}^\dagger \frac{c_{2'\uparrow}^\dagger - c_{1'\uparrow}^\dagger}{\sqrt{2}}\ket{\Omega}$.
Both states possess energy $E_m^{(0)} = -t$ with respective weights $-\sqrt{2}t'/3$ and $-2\sqrt{2}t'/3$, which yields the second-order energy correction:
\begin{equation}
    \Delta E_q^{(2)} = -\frac{10{t'}^2}{27t}
\end{equation}
Notably, the $t'$ sign dependence vanishes due to quadratic terms in the perturbation expansion.

Similar calculations for $S=1$ and $S=0$ channels produce:
\begin{equation}
\begin{aligned}
    H^\prime\ket{S=1}=-\frac{t^\prime}{6}&\left(-2\sqrt{2}c_{1\uparrow}^\dagger c_{2\uparrow}^\dagger c_{3\downarrow}^\dagger\frac{c_{1^\prime\uparrow}^\dagger-c_{2^\prime\uparrow}^\dagger}{\sqrt{2}}\right.\\
    &\left.+2\sqrt{2}c_{1\uparrow}^\dagger c_{2\uparrow}^\dagger c_{3\uparrow}^\dagger\frac{c_{1^\prime\downarrow}^\dagger-c_{2^\prime\downarrow}^\dagger}{\sqrt{2}}-\sqrt{2}\frac{c_{1\downarrow}^\dagger-c_{2\downarrow}^\dagger}{\sqrt{2}}c_{1^\prime\uparrow}^\dagger c_{2^\prime\uparrow}^\dagger c_{3^\prime\uparrow}^\dagger\right.\\
    &\left.+\sqrt{6}\frac{c_{1\uparrow}^\dagger-c_{2\uparrow}^\dagger}{\sqrt{2}}\frac{c_{1^\prime\downarrow}^\dagger c_{2^\prime\uparrow}^\dagger c_{3^\prime\uparrow}^\dagger+c_{1^\prime\uparrow}^\dagger c_{2^\prime\downarrow}^\dagger c_{3^\prime\uparrow}^\dagger-c_{1^\prime\uparrow}^\dagger c_{2^\prime\uparrow}^\dagger c_{3^\prime\downarrow}^\dagger}{\sqrt{3}}\right)\ket{\Omega}\\
    H^\prime\ket{S=0}=-\frac{t^\prime}{6\sqrt{3}}&\left(2\sqrt{3}\frac{c_{1\uparrow}^\dagger-c_{2\uparrow}^\dagger}{\sqrt{2}}\frac{2c_{1^\prime\downarrow}^\dagger c_{2^\prime\downarrow}^\dagger c_{3^\prime \uparrow}^\dagger-c_{1^\prime\uparrow}^\dagger c_{2^\prime\downarrow}^\dagger c_{3^\prime \downarrow}^\dagger-c_{1^\prime\downarrow}^\dagger c_{2^\prime\uparrow}^\dagger c_{3^\prime \downarrow}^\dagger}{\sqrt{3}}\right.\\
    &+2\sqrt{3}\frac{c_{1\downarrow}^\dagger-c_{2\downarrow}^\dagger}{\sqrt{2}}\frac{2c_{1^\prime\uparrow}^\dagger c_{2^\prime\uparrow}^\dagger c_{3^\prime \downarrow}^\dagger-c_{1^\prime\uparrow}^\dagger c_{2^\prime\downarrow}^\dagger c_{3^\prime \uparrow}^\dagger-c_{1^\prime\downarrow}^\dagger c_{2^\prime\uparrow}^\dagger c_{3^\prime \uparrow}^\dagger}{\sqrt{3}}\\
    &+2\sqrt{3}\frac{c_{1\uparrow}^\dagger c_{2\downarrow}^\dagger c_{3\downarrow}^\dagger+c_{1\downarrow}^\dagger c_{2\uparrow}^\dagger c_{3\downarrow}^\dagger-2c_{1^\downarrow}^\dagger c_{2\downarrow}^\dagger c_{3\uparrow}^\dagger}{\sqrt{6}}\frac{c_{1^\prime\uparrow}^\dagger -c_{2^\prime\uparrow}^\dagger }{\sqrt{2}}\\
    &\left.+2\sqrt{3}\frac{c_{1\downarrow}^\dagger c_{2\uparrow}^\dagger c_{3\uparrow}^\dagger+c_{1\uparrow}^\dagger c_{2\downarrow}^\dagger c_{3\uparrow}^\dagger-2c_{1\uparrow}^\dagger c_{2\uparrow}^\dagger c_{3\downarrow}^\dagger}{\sqrt{6}}\frac{c_{1^\prime\downarrow}^\dagger-c_{2^\prime\downarrow}^\dagger}{\sqrt{2}}\right)\ket{\Omega}\\
\end{aligned}
\end{equation}
which result in the following energy corrections:
\begin{equation}
\begin{aligned}
    \Delta E_t^{(2)} &= -\frac{6{t'}^2}{27t} \\
    \Delta E_s^{(2)} &= -\frac{4{t'}^2}{27t}
\end{aligned}
\end{equation}

This energy hierarchy ($J, -J, -2J$) corresponds to quintet ($S=2$), triplet ($S=1$), and singlet ($S=0$) couplings, and it reveals the effective exchange constant:
\begin{equation}
    J = -\frac{2{t'}^2}{27t}
\end{equation}

\begin{figure}[htp]
    \centering
    \includegraphics[width=0.5\linewidth]{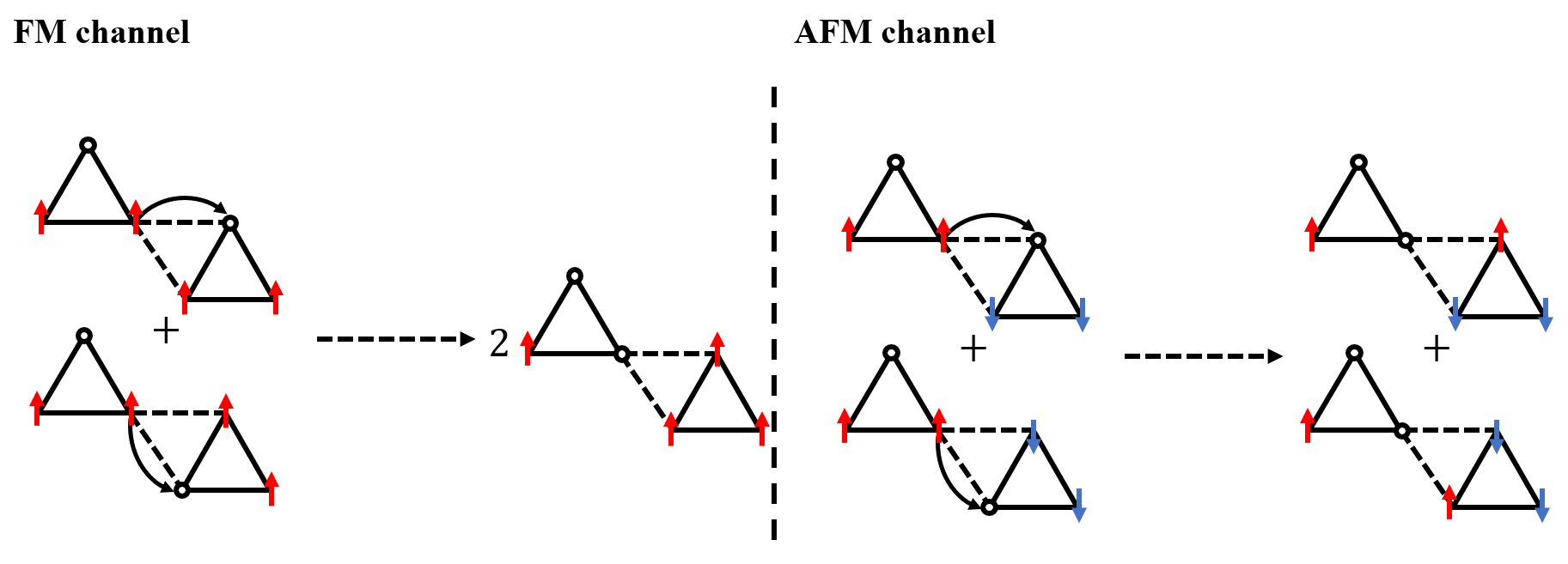}
    \caption{The left panel demonstrates that two configurations within the ferromagnetic channel generate identical virtual intermediate states, thereby enhancing quantum coherence through constructive interference; 
    conversely, analogous configurations in antiferromagnetic channels access distinct final states despite equivalent hopping processes. When combined with the energetic degeneracy of all intermediate states, this disparity creates an interference enhancement mechanism analogous to the inequality $2^2 > 1^2 + 1^2$, which fundamentally establishes ferromagnetic coupling as the dominant interaction mechanism.}
    \label{fig:comparison_coherent}
\end{figure}

The emergence of ferromagnetism in the $U\to\infty$ regime under the condition $|t'|\ll t$ originates from fundamentally quantum interference mechanisms: 
the fully polarized $S=2$ channel exhibits constructive interference characteristics, wherein all virtual hopping processes coherently interfere as they lead to the same state, whereas antiferromagnetic configurations exhibit diminished quantum coherence, arising from competing spin orientation channels within three-electron trimer systems. 
A schematic diagram elucidating this mechanism is presented in Fig.~\ref{fig:comparison_coherent}.

This analytical framework remains applicable to subsequent calculations.


\subsection{Effect of a staggered flux $|\phi|<\pi/2$}
\label{App:resistance to flux}

\begin{figure}[t]
\centering
\includegraphics[width=0.3\columnwidth]{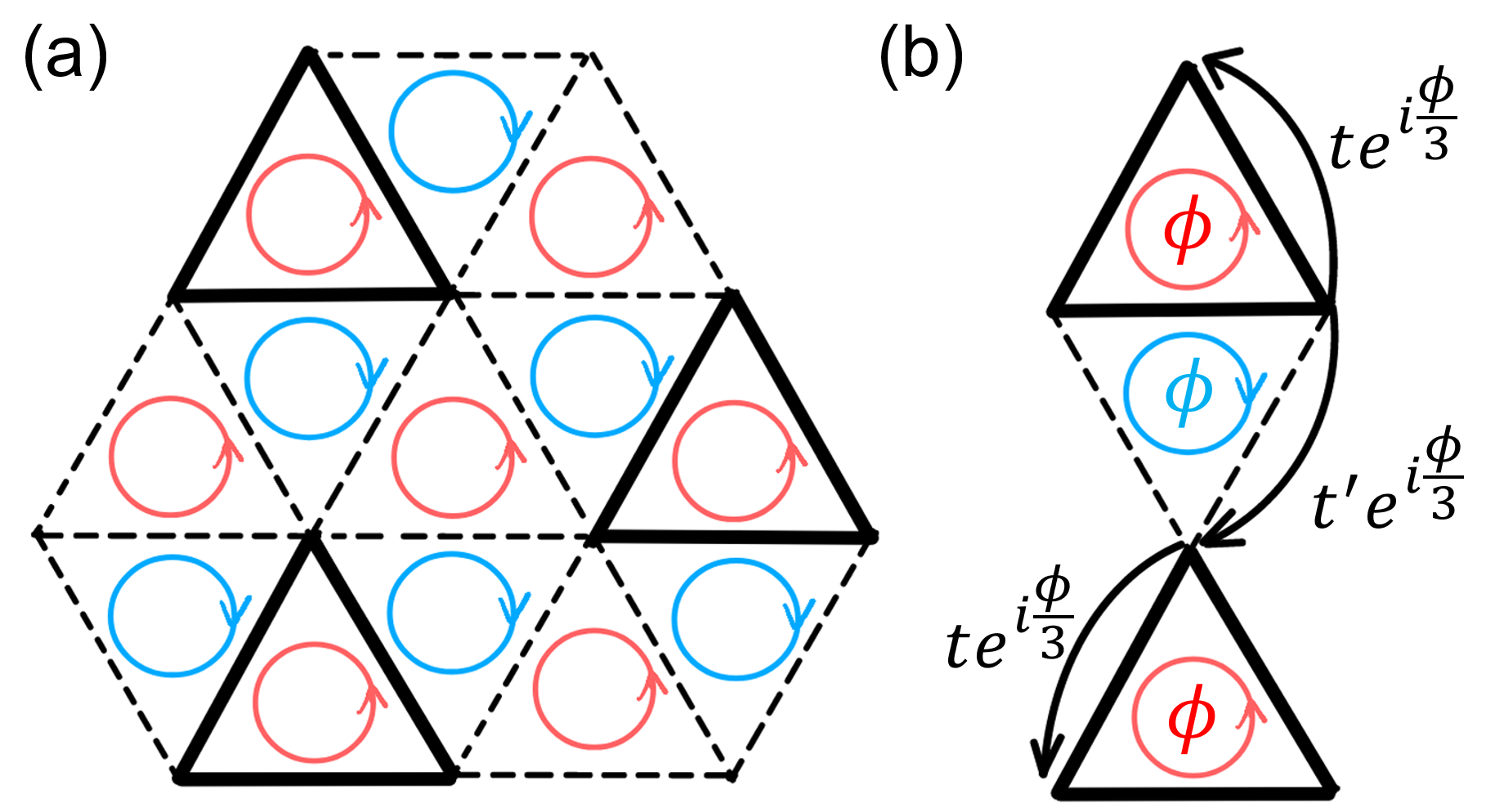}
\caption{(a) The trimerized triangular lattice with a staggered flux pattern $\pm\phi$. 
(b) The phase convention of hopping integrals.}
\label{fig:flux}
\end{figure}

When a staggered flux $\phi$ threads each trimer, the spectrum undergoes a shift and the intra- and inter-trimer hopping amplitudes carry a phase, as shown in Fig.~\ref{fig:flux}.
Nonetheless, the ferromagnetic structure remains robust, which will be justified below.
In addition, the conclusion holds regardless of the sign of flux, as the ground state remains unchanged before $|\phi|$ reaches $\frac{\pi}{2}$.
In the discussion below, we focus on positive $\phi$, but the conclusion applies equally to negative $\phi$.

Now, the ground-state energy of these channels is modified as $-4t\cos{\theta}$.
All intermediate states are calculated below:
\begin{equation}
\begin{aligned}
    H^\prime\ket{S=2}=\frac{2t^\prime}{3}i&\left[\left(\sin(\theta+\frac{\pi}{6})\ket{q_{-1}}_A+\sin(\theta-\frac{\pi}{6})\ket{q_{1}}_A\right)c_{1^\prime\uparrow}^\dagger c_{2^\prime\uparrow}^\dagger c_{3^\prime\uparrow}^\dagger\right.\\
    &\left.+\cos{\theta}c_{1\uparrow}^\dagger c_{2\uparrow}^\dagger c_{3\uparrow}^\dagger\left(\ket{q_{-1}}_B-\ket{q_{1}}_B\right)\right]\ket{\Omega}\\
    H^\prime\ket{S=1}=\frac{t^\prime}{3}&i\left[\left(\sin(\theta+\frac{\pi}{6})\ket{q_{-1}}_A+\sin(\theta-\frac{\pi}{6})\ket{q_{1}}_A\right)(c_{1^\prime\uparrow}^\dagger c_{2^\prime\downarrow}^\dagger c_{3^\prime\uparrow}^\dagger+c_{1^\prime\downarrow}^\dagger c_{2^\prime\uparrow}^\dagger c_{3^\prime\uparrow}^\dagger-c_{1^\prime\uparrow}^\dagger c_{2^\prime\uparrow}^\dagger c_{3^\prime\downarrow}^\dagger)\right.\\
    &\left.-\left(\sin(\theta+\frac{\pi}{6})\ket{q_{-1}}_A+\sin(\theta-\frac{\pi}{6})\ket{q_{1}}_A\right)c_{1^\prime\uparrow}^\dagger c_{2^\prime\uparrow}^\dagger c_{3^\prime\uparrow}^\dagger\right.\\
    &\left.+\cos{\theta}c_{1\uparrow}^\dagger c_{2\uparrow}^\dagger c_{3\uparrow}^\dagger\left(\ket{q_{-1}}_B-\ket{q_{1}}_B\right)\right.\\
    &\left.+(-\sin\theta c_{1\uparrow}^\dagger c_{2\uparrow}^\dagger c_{3\downarrow}^\dagger+\sin\theta c_{1\downarrow}^\dagger c_{2\uparrow}^\dagger c_{3\uparrow}^\dagger-i\cos\theta c_{1\uparrow}^\dagger c_{2\uparrow}^\dagger c_{3\downarrow}^\dagger)\left(\ket{q_{-1}}_B-\ket{q_{1}}_B\right)\right]\ket{\Omega}\\
    H^\prime\ket{S=0}=\frac{t^\prime}{6\sqrt{3}}&i\left[2\left(\sin(\theta+\frac{\pi}{6})\ket{q_{-1}}_A+\sin(\theta-\frac{\pi}{6})\ket{q_{1}}_A\right)(2c_{1^\prime\downarrow}^\dagger c_{2^\prime\downarrow}^\dagger c_{2^\prime\uparrow}^\dagger-c_{1^\prime\uparrow}^\dagger c_{2^\prime\uparrow}^\dagger c_{2^\prime\downarrow}^\dagger-c_{1^\prime\downarrow}^\dagger c_{2^\prime\uparrow}^\dagger c_{2^\prime\downarrow}^\dagger)\right.\\
    &\left.+2\left(\sin(\theta+\frac{\pi}{6})\ket{q_{-1}}_A+\sin(\theta-\frac{\pi}{6})\ket{q_{1}}_A\right)(2c_{1^\prime\uparrow}^\dagger c_{2^\prime\uparrow}^\dagger c_{3^\prime\downarrow}^\dagger-c_{1^\prime\uparrow}^\dagger c_{2^\prime\downarrow}^\dagger c_{3^\prime\uparrow}^\dagger-c_{1^\prime\downarrow}^\dagger c_{2^\prime\uparrow}^\dagger c_{3^\prime\uparrow}^\dagger)\right.\\
    &\left.+\left((2e^{-i\theta}-e^{i\theta})c_{1\downarrow}^\dagger c_{2\uparrow}^\dagger c_{3\uparrow}^\dagger+(2e^{i\theta}-e^{-i\theta})c_{1\uparrow}^\dagger c_{2\downarrow}^\dagger c_{3\uparrow}^\dagger-2\cos\theta c_{1\uparrow}^\dagger c_{2\uparrow}^\dagger c_{3\downarrow}^\dagger\right)\left(\ket{q_{-1}}_B-\ket{q_{1}}_B\right)\right.\\
    &\left.+\left((2e^{-i\theta}-e^{i\theta})c_{1\uparrow}^\dagger c_{2\downarrow}^\dagger c_{3\downarrow}^\dagger+(2e^{i\theta}-e^{-i\theta})c_{1\downarrow}^\dagger c_{2\uparrow}^\dagger c_{3\downarrow}^\dagger-2\cos\theta c_{1\downarrow}^\dagger c_{2\downarrow}^\dagger c_{3\uparrow}^\dagger\right)\left(\ket{q_{-1}}_B-\ket{q_{1}}_B\right)\right]\ket{\Omega}\\
\end{aligned}
\end{equation}
where $\ket{q_{-1}}$ and $\ket{q_{1}}$ are the eigenstate solved in SM.~\ref{App:eigenstate of trimer at n=1}, with related eigenenergy $2t\cos(\theta-2\pi/3)$ and $2t\cos(\theta+2\pi/3)$, and $c_{1\sigma_1}^\dagger c_{2\sigma_2}^\dagger c_{3\sigma_3}^\dagger\ket{\Omega}$ is already the eigenstate with an eigenenergy $0$ under the condition $U\rightarrow\infty$.

The unperturbed energy is $-4t\cos{\theta}$.
Therefore, the corresponding energy gains are
\begin{equation}
\begin{aligned}
    &\Delta E_q^{(2)}=-\frac{2t^{\prime 2}}{9t}\left(\frac{\sin^2{(\theta+\pi/6)}+\cos^2{\theta}}{\sqrt{3}\cos{(\theta-\pi/6)}}+\frac{\sin^2{(\theta-\pi/6)}+\cos^2{\theta}}{\sqrt{3}\cos{(\theta+\pi/6)}}\right)\\
    &\Delta E_t^{(2)}=-\frac{2t^{\prime 2}}{9t}\left(\frac{\sin^2{(\theta+\pi/6)+1/2}}{\sqrt{3}\cos{(\theta-\pi/6)}}+\frac{\sin^2{(\theta-\pi/6)+1/2}}{\sqrt{3}\cos{(\theta+\pi/6)}}\right)\\
    &\Delta E_s^{(2)}=-\frac{2t^{\prime 2}}{9t}\left(\frac{\sin^2{(\theta+\pi/6)}+\frac{1}{4}(3-2\cos^2{\theta})}{\sqrt{3}\cos{(\theta-\pi/6)}}+\frac{\sin^2{(\theta-\pi/6)}+\frac{1}{4}(3-2\cos^2{\theta})}{\sqrt{3}\cos{(\theta+\pi/6)}}\right)\\
\end{aligned}
\label{SMeq: energy gain with flux}
\end{equation}
where $\theta=\phi/3$ has been used here to maintain the convention in main text. 
The energy shifts maintain the ratio characteristic of a spin-1 Heisenberg model, specifically $\Delta E\propto S(S+1)$
The effective coupling with flux $\phi$ reads
\begin{equation}
\begin{aligned}
    J(\phi)&=-\frac{{t^\prime}^2}{18t}\frac{\cos\frac{2\phi}{3} \cos\frac{\phi}{3}}{\cos(\frac{\phi}{3}+\frac{\pi}{6})  \cos(\frac{\phi}{3}-\frac{\pi}{6})}\\
    \approx&J\left(1-\frac{7}{54}\phi^2\right)
\end{aligned}
\end{equation}
It indicates that threading a flux slightly renormalizes the effective coupling $J$, reducing its magnitude without changing the sign.

\subsection{A staggered flux $|\phi|=\pi/2$ is threading}

\label{App:energy with flux}

When $|\phi|=\frac{\pi}{2}$, singlet $\ket{k_{-1}}$ and triplet $\ket{k_0}$ states become degenerate ground states of one trimer, with energy $-\sqrt{3}t$. 
For the channels between two trimers both in triplet states, the energy gain can be directly obtained by setting $|\phi|=\pi/2$ according to Eq.~\eqref{SMeq: energy gain with flux}.
We must further calculate the energy corrections between trimers where one or both are in singlet states, as shown below (for simplicity, take $\phi=\pi/2$ case for an example):

\begin{equation}
\begin{aligned}
    H^\prime\ket{k_0}_A\ket{k_{-1}}_B=
    \frac{t^\prime}{3\sqrt{2}}&\left[\sqrt{3}e^{i\frac{\pi}{6}}\ket{q_{-1}}_A(c_{1^\prime\uparrow}^\dagger c_{2^\prime\downarrow}^\dagger c_{3^\prime\uparrow}^\dagger-c_{1^\prime\downarrow}^\dagger c_{2^\prime\uparrow}^\dagger c_{3^\prime\uparrow}^\dagger)\right.\\
    &\left.-(c_{1\downarrow}^\dagger c_{2\uparrow}^\dagger c_{3\uparrow}^\dagger+e^{i\frac{\pi}{3}}c_{1\uparrow}^\dagger c_{2\downarrow}^\dagger c_{3\uparrow}^\dagger)\cdot e^{i\frac{2\pi}{3}}\left(\ket{q_{-1}}_B-\ket{q_{1}}_B\right)\right.\\
    &\left.+\sqrt{3}e^{i\frac{\pi}{6}}c_{1\uparrow}^\dagger c_{2\uparrow}^\dagger c_{3\uparrow}^\dagger\cdot e^{i\frac{2\pi}{3}}\left(\ket{q_{-1}}_B-\ket{q_{1}}_B\right)\right]\ket{\Omega}\\
    H^\prime\ket{k_{-1}}_A\ket{k_0}_B=\frac{t^\prime}{3\sqrt{2}}&\left[-e^{-i\frac{\pi}{3}}\left(\ket{q_{-1}}_A+2\ket{q_{1}}_A\right)c_{1^\prime\uparrow}^\dagger c_{2^\prime\uparrow}^\dagger c_{3^\prime\downarrow}^\dagger\right.\\
    &\left.+e^{-i\frac{\pi}{3}}\left(\ket{q_{-1}}_A+2\ket{q_{1}}_A\right)c_{1\prime\uparrow}^\dagger c_{2\prime\uparrow}^\dagger c_{3\prime\uparrow}^\dagger\right.\\
    &\left.+(-e^{i\frac{\pi}{6}}c_{1\uparrow}^\dagger c_{2\uparrow}^\dagger c_{3\downarrow}^\dagger+e^{-i\frac{\pi}{6}}c_{1\uparrow}^\dagger c_{2\downarrow}^\dagger c_{3\uparrow}^\dagger+e^{i\frac{\pi}{2}}c_{1\downarrow}^\dagger c_{2\uparrow}^\dagger c_{3\uparrow}^\dagger)\cdot i\left(\ket{q_{-1}}_B-\ket{q_{1}}_B\right)\right]\ket{\Omega}\\
    H^\prime\ket{k_{-1}}_A\ket{k_{-1}}_B=\frac{t^\prime}{6}&\left[-e^{i\frac{\pi}{3}}\left(\ket{q_{-1}}_A+2\ket{q_{1}}_A\right)(c_{1^\prime\uparrow}^\dagger c_{2^\prime\downarrow}^\dagger c_{3^\prime\uparrow}^\dagger-c_{1^\prime\downarrow}^\dagger c_{2^\prime\uparrow}^\dagger c_{3^\prime\uparrow}^\dagger)\right.\\
    &\left.+e^{i\frac{\pi}{3}}\left(\ket{q_{-1}}_A+2\ket{q_{1}}_A\right)(c_{1^\prime\uparrow}^\dagger c_{2^\prime\downarrow}^\dagger c_{3^\prime\downarrow}^\dagger-c_{1^\prime\downarrow}^\dagger c_{2^\prime\uparrow}^\dagger c_{3^\prime\downarrow}^\dagger)\right.\\
    &\left.+(e^{-i\frac{\pi}{6}}c_{1\uparrow}^\dagger c_{2\downarrow}^\dagger c_{3\uparrow}^\dagger+e^{i\frac{\pi}{2}}c_{1\downarrow}^\dagger c_{2\uparrow}^\dagger c_{3\uparrow}^\dagger-e^{i\frac{\pi}{6}}c_{1\uparrow}^\dagger c_{2\uparrow}^\dagger c_{3\downarrow}^\dagger)\frac{e^{-i\frac{\pi}{3}}}{\sqrt{3}}\left(-\ket{q_{-1}}_B+2\ket{q_{1}}_B-\ket{q_{0}}_B\right)\right.\\
    &\left.+(e^{-i\frac{\pi}{6}}c_{1\downarrow}^\dagger c_{2\uparrow}^\dagger c_{3\downarrow}^\dagger-e^{i\frac{\pi}{6}}c_{1\downarrow}^\dagger c_{2\downarrow}^\dagger c_{3\uparrow}^\dagger+e^{i\frac{\pi}{2}}c_{1\uparrow}^\dagger c_{2\downarrow}^\dagger c_{3\downarrow}^\dagger)\frac{e^{-i\frac{\pi}{3}}}{\sqrt{3}}\left(-\ket{q_{-1}}_B+2\ket{q_{1}}_B-\ket{q_{0}}_B\right)\right]\ket{\Omega}\\
\end{aligned}
\end{equation}
where $\ket{q_{-1}},\ \ket{q_{1}},\ \ket{q_{0}}$ are eigenstates, with an eigenenergies $0,-\sqrt{3}t,\sqrt{3}t$. 
As the energy of fully occupancy states $c_{1\sigma_1}^\dagger c_{2\sigma_2}^\dagger c_{3\sigma_3}^\dagger\ket{\Omega}$ is $0$ for $U\rightarrow\infty$, the total energy of virtual intermediate states are $0,-\sqrt{3}t,\sqrt{3}t$.
Combined with the unperturbed energy is $-2\sqrt{3}t$.
The above virtual excited state induces the energy gain as $-\frac{7t^{\prime 2}}{12\sqrt{3}t},-\frac{3t^{\prime 2}}{4\sqrt{3}t},-\frac{83{t^\prime}^2}{108\sqrt{3}t}$.

Unlike the former calculations, different channels are separated by symmetry, and the mixture between different channels is naturally forbidden.
Here, according to the decomposition rule of the SU(2) representations, the two trimers have 6 spin channels, which are separated into three sectors.
The quintet sector does not mix with others with energy 
$\Delta E^{(2)}_q=-\frac{2{t^\prime}^2}{3\sqrt{3}t}$.
The two belonging to the singlet sector could mix in principle; nevertheless, a straightforward calculation shows that the mixing matrix elements vanish at the 2nd order. 
The energies are simply the diagonal term 
$\Delta E^{(2)}_{s,1}=-\frac{5 {t^\prime}^2}{12\sqrt{3}t} $
and $\Delta E^{(2)}_{s,2}=
-\frac{83{t^\prime}^2}{108\sqrt{3}t}$, which lies on the two ends of the energy spectrum.
 For the three triplet sector, their mixing matrix elements are calculated 
as:
\begin{equation}
    {H_{ab}^\prime}^{(2)}=\sum_{m}\frac{\bra{\psi_{a,s_z}}H^\prime\ket{m}\bra{m}H^\prime\ket{\psi_{b,s_z} }}{E^{(0)}-E_{m}^{(0)}},
    \label{Eq:off-diagonal}
\end{equation}
where $a,b$ are the indices of the triplet sector; $s_z$ can take any value among $\pm 1, 0$ yielding the same matrix elements according to the Wigner-Eckart theorem. 
Consequently, the secular equation reads
\begin{equation}
\Delta H^{(2)}_{S_{\mathrm{tot}}=1}/\left(\frac{t'^2}{t}\right)=
\begin{pmatrix}
-\frac{1}{2\sqrt{3}}&-\frac{e^{i\pi/3}}{12\sqrt{6}}&0\\
-\frac{e^{-i\pi/3}}{12\sqrt{6}}&-\frac{7}{12\sqrt{3}}&0\\
0&0&-\frac{3}{4\sqrt{3}} \\
\end{pmatrix}
\end{equation}
yielding the eigenenergies 
$-\frac{13\pm \sqrt{3}}{24\sqrt{3}} 
\frac{t'^2}{t}$
and $-\frac{3}{4\sqrt{3}}\frac{t'^2}{t}$. 

The above result shows that when threading flux makes the singlet and triplet degenerate in the decoupling limit, the system is non-magnetic (all trimers stay in the singlet state $\ket{k_{-1}}$) rather than FM due to the second-order perturbation from $t'$ coupling.

\section{Phase transitions by tuning $t^\prime$ and $U$}

\subsection{Phase transition in the trimerized triangular lattice}

\label{App:CompetingAFM&FM}

In the preceding content, we established that in the trimerized triangular lattice, FM is dominant when double occupancy states are completely suppressed, because the intermediate states of the FM channel are more coherent than those of AFM channel. 
However, when the interaction becomes finite, the occurrence of double occupancy intermediate states opens more channels to further reduce the energy for AFM coupling, resulting in the competition between the FM phase and AFM phase.
We now elaborate on the calculations, which include the contributions of double occupancy states.
For simplicity, the following discussion only focuses on the case where there is no flux.

The calculation of state $\ket{S=2}$ is the same as our previous derivation, because intermediate states with double occupancy are forbidden by the Pauli exclusion principle in the fully polarized sector. 
The calculation for triplet and singlet sectors containing double occupancy is shown below:
\begin{equation}
\begin{aligned}
    H^\prime\ket{S=1}=\frac{\sqrt{2}t^\prime}{6}&\left[\frac{c_{1\uparrow}^\dagger-c_{2\uparrow}^\dagger}{\sqrt{2}}\left(\frac{1}{\sqrt{3}}(\ket{\psi_1,0}_B-2\ket{\psi_1,\frac{2\pi}{3}}_B-2\ket{\psi_1,-\frac{2\pi}{3}}_B)-2c_{3^\prime\uparrow}^\dagger c_{3^\prime\downarrow}^\dagger c_{1^\prime\uparrow}^\dagger -2c_{3^\prime\uparrow}^\dagger c_{3^\prime\downarrow}^\dagger c_{2^\prime\uparrow}^\dagger \right)\right.\\
    &\left.-\frac{c_{1\downarrow}^\dagger-c_{2\downarrow}^\dagger}{\sqrt{2}}c_{1^\prime\uparrow}^\dagger c_{2^\prime\uparrow}^\dagger c_{3^\prime\uparrow}^\dagger+2c_{1\uparrow}^\dagger c_{2\uparrow}^\dagger c_{3\uparrow}^\dagger\frac{c_{1^\prime\downarrow}^\dagger-c_{2^\prime\downarrow}^\dagger}{\sqrt{2}}\right.\\
    &\left.-2\left(\frac{1}{\sqrt{3}}(\ket{\psi_1,0}_A+\ket{\psi_1,\frac{2\pi}{3}}_A+\ket{\psi_1,-\frac{2\pi}{3}}_A)+c_{1\uparrow}^\dagger c_{1\downarrow}^\dagger c_{2\uparrow}^\dagger+c_{2\uparrow}^\dagger c_{2\downarrow}^\dagger c_{3\uparrow}^\dagger\right.\right.\\
    &\left.\left.-c_{1\uparrow}^\dagger c_{1\downarrow}^\dagger c_{3\uparrow}^\dagger-c_{2\uparrow}^\dagger c_{2\downarrow}^\dagger c_{1\uparrow}^\dagger \right)\frac{c_{1^\prime\uparrow}^\dagger-c_{2^\prime\uparrow}^\dagger}{\sqrt{2}}\right]\ket{\Omega}\\
    H^\prime\ket{S=0}=\frac{t^\prime}{3\sqrt{2}}&\left[\frac{c_{1\uparrow}^\dagger-c_{2\uparrow}^\dagger}{\sqrt{2}}\left(\sqrt{3}\ket{\psi_1,\frac{2\pi}{3}}_B+\sqrt{3}\ket{\psi_1,-\frac{2\pi}{3}}_B+3c_{3^\prime\uparrow}^\dagger c_{3^\prime\downarrow}^\dagger c_{1^\prime\downarrow}^\dagger-3c_{3^\prime\uparrow}^\dagger c_{3^\prime\downarrow}^\dagger c_{2^\prime\downarrow}^\dagger\right)\right.\\
    &\left.+\frac{c_{1\downarrow}^\dagger-c_{2\downarrow}^\dagger}{\sqrt{2}}\left(\sqrt{3}\ket{\psi_1,\frac{2\pi}{3}}_B+\sqrt{3}\ket{\psi_1,-\frac{2\pi}{3}}_B-3c_{3^\prime\uparrow}^\dagger c_{3^\prime\downarrow}^\dagger c_{1^\prime\uparrow}^\dagger+3c_{3^\prime\uparrow}^\dagger c_{3^\prime\downarrow}^\dagger c_{2^\prime\uparrow}^\dagger\right)\right.\\
    &\left.-\left(\sqrt{3}\ket{\psi_1,\frac{2\pi}{3}}_A+\sqrt{3}\ket{\psi_1,-\frac{2\pi}{3}}_A+3c_{1\uparrow}^\dagger c_{1\downarrow}^\dagger c_{2\uparrow}^\dagger+3c_{2\uparrow}^\dagger c_{2\downarrow}^\dagger c_{3\uparrow}^\dagger-3c_{1\uparrow}^\dagger c_{1\downarrow}^\dagger c_{3\uparrow}^\dagger-3c_{2\uparrow}^\dagger c_{2\downarrow}^\dagger c_{1\uparrow}^\dagger\right)\frac{c_{1^\prime\downarrow}^\dagger-c_{2^\prime\downarrow}^\dagger}{\sqrt{2}}\right.\\
    &\left.-\left(\sqrt{3}\ket{\psi_1,\frac{2\pi}{3}}_A+\sqrt{3}\ket{\psi_1,-\frac{2\pi}{3}}_A-3c_{1\uparrow}^\dagger c_{1\downarrow}^\dagger c_{2\downarrow}^\dagger-3c_{2\uparrow}^\dagger c_{2\downarrow}^\dagger c_{3\downarrow}^\dagger+3c_{1\uparrow}^\dagger c_{1\downarrow}^\dagger c_{3\downarrow}^\dagger+3c_{2\uparrow}^\dagger c_{2\downarrow}^\dagger c_{1\downarrow}^\dagger\right)\frac{c_{1^\prime\uparrow}^\dagger-c_{2^\prime\uparrow}^\dagger}{\sqrt{2}}\right]\\
\end{aligned}
\end{equation}
where virtual intermediate states consist of trimer pairs with (1,3) electron occupations:  $ \frac{c_{1\sigma}^\dagger-c_{2\sigma}^\dagger}{\sqrt{2}}\ket{\Omega}$ are the trimer filled with 1 electron with energy $-t$, and $\ket{\psi_1,0},\ \ket{\psi_1,\pm\frac{2\pi}{3}},\ c_{i\uparrow}^\dagger c_{i\downarrow}^\dagger c_{j\sigma_j}^\dagger\ket{\Omega}$ are the ones filled with 3 electrons with energies $0,-6t^2/U, U$.
The unperturbed energy is $-4t$.
The calculation shows the corresponding energy gains are approximately:
\begin{equation}
\begin{aligned}
    \Delta E_q^{(2)}&=-\frac{10t^{\prime 2}}{27t}\\
    \Delta E_t^{(2)}&=-\left(\frac{6t^{\prime 2}}{27t}+\frac{124t^{\prime 2}}{81U}\right)\\
    \Delta E_s^{(2)}&=-\left(\frac{4t^{\prime 2}}{27t}+\frac{186t^{\prime 2}}{81U}\right)\\
\end{aligned}
\end{equation}
which yields an effective coupling $J$, comprising both ferromagnetic and anti-ferromagnetic superexchanges:
\begin{equation}
    J=-\frac{2t^{\prime 2}}{27t}+\frac{62t^{\prime 2}}{81U}
    \label{eq:FM_and_AFM_coefficient}
\end{equation}
The opposite sign of $t$- and $U$-term illustrates the competition between FM and AFM.
Eventually, by setting Eq.~\eqref{eq:FM_and_AFM_coefficient} to zero, the critical transition interaction for the FM-AFM transition is determined to be $U_c/t=31/3\approx 10.3$ at the level of 2nd-order perturbation theory.

\begin{figure}[t]
    \centering
    \subfigure[ED result]{
    \label{fig:transition U}
    \includegraphics[width=0.33\columnwidth]{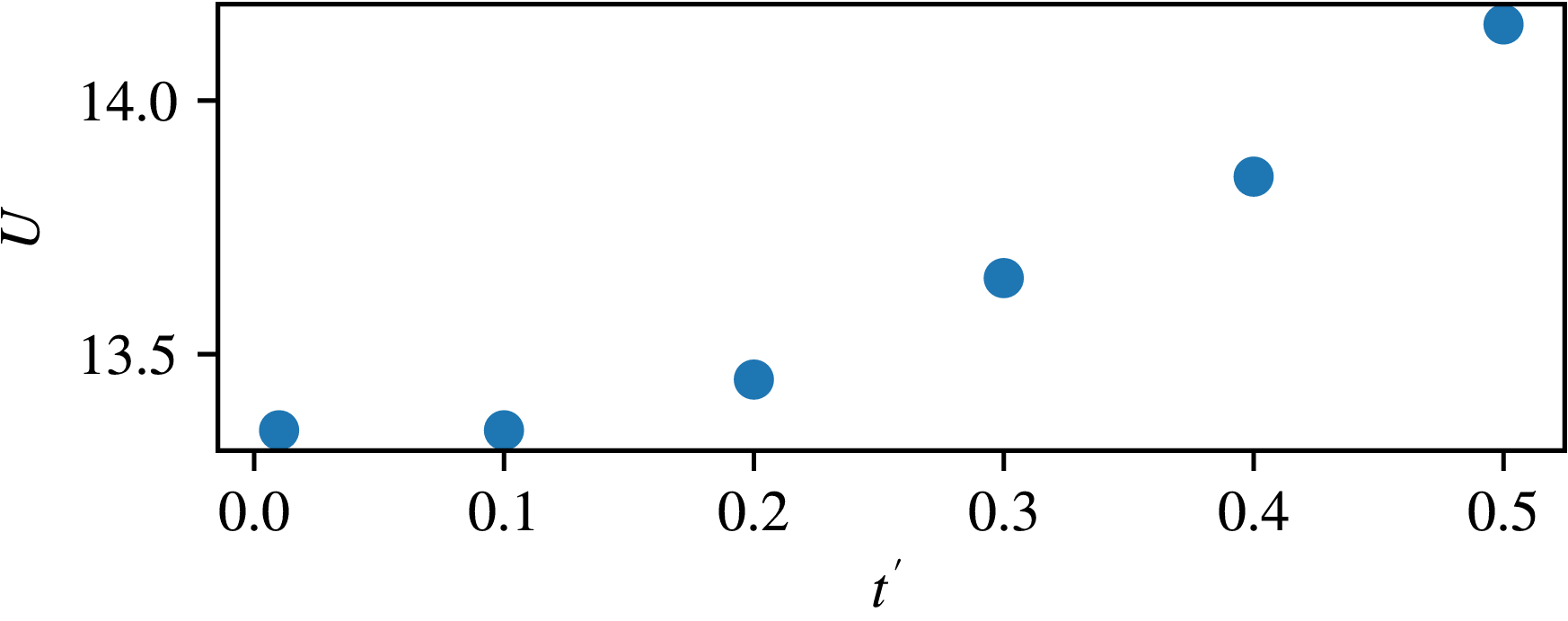}}\quad
    \subfigure[Higher order hopping in AFM channel]{
    \label{fig:AFM coefficient}
    \includegraphics[width=0.33\columnwidth]{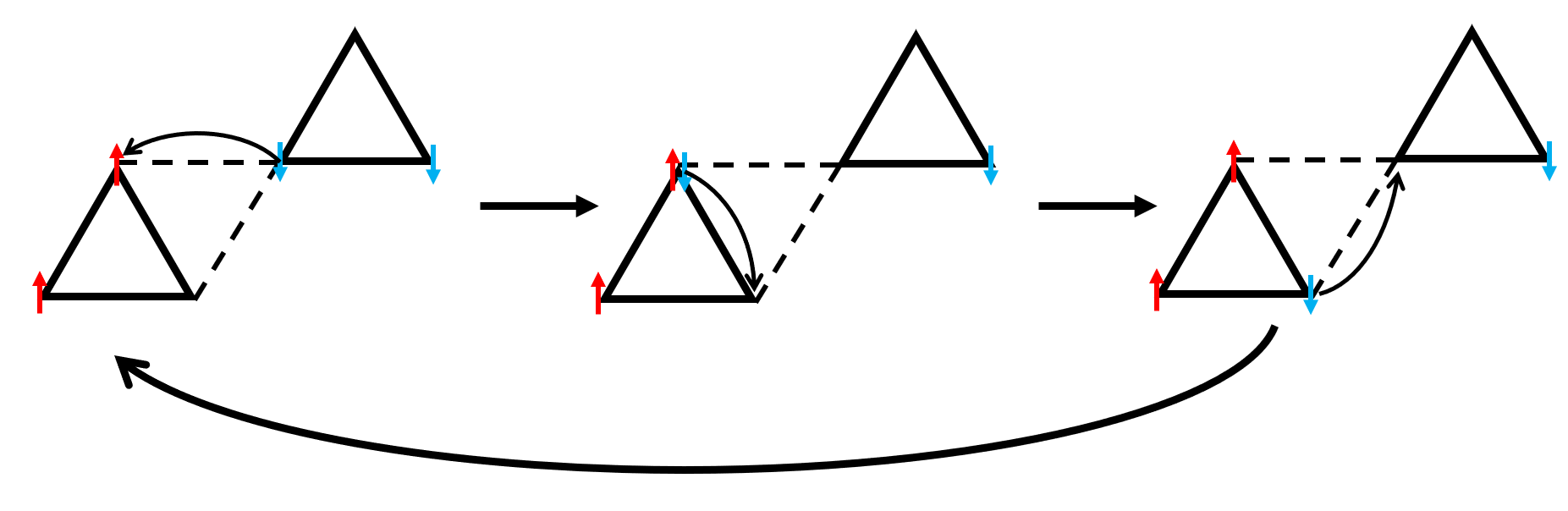}}\quad
    \caption{(a)Two-cluster estimation of spin exchange: The dependence of transition $U$ between FM\&AFM with respect to $t^\prime$. Here $t=1$. 
    (b) An example of a higher order process included in ED results.
    The energy correction needs 3rd-order perturbation theory, but is at the order of $t^{\prime 2}/U$. }
\end{figure}

\subsection{FM/AFM competition in the trimerized Kagome lattice}
\label{App:kagome AFM&FM}

A similar calculation can be easily generalized to the trimerized Kagome lattice. 

\begin{figure}[ht]
    \centering
    \subfigure[$t^\prime/t=0.2$]{\includegraphics[width=0.2\columnwidth]{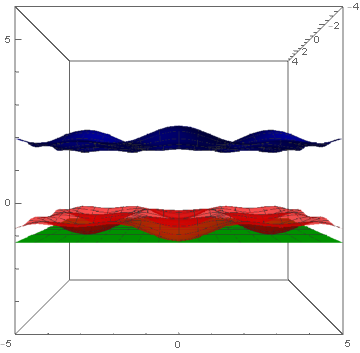}}
    \subfigure[$t^\prime/t=0.8$]{\includegraphics[width=0.2\columnwidth]{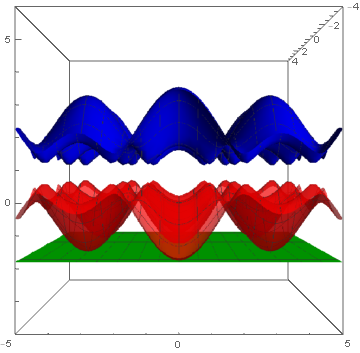}}\\
    \caption{The free band structure of trimerized Kagome lattice with respect to ratio $t^\prime/t$. The lowest band (green one) is flat regardless of the ratio $t^\prime/t$}
    \label{Fig:Kagome spectrum}
\end{figure}

We begin with the Bloch Hamiltonian for the trimerized Kagome lattice:
\begin{equation}
\begin{pmatrix}
    0&t+t^\prime e^{-i\vec{k}\cdot\vec{a}_1}&t+t^\prime e^{-i\vec{k}\cdot\vec{a}_2}\\
    t+t^\prime e^{i\vec{k}\cdot\vec{a}_1}&0&t+t^\prime e^{-i\vec{k}\cdot(\vec{a}_1+\vec{a}_2)}\\
    t+t^\prime e^{i\vec{k}\cdot\vec{a}_2}&t+t^\prime e^{i\vec{k}\cdot(\vec{a}_1+\vec{a}_2)}&0\\
\end{pmatrix}
\end{equation}
where $\vec{k}=(k_x,k_y)$ and $\vec{a}_1=(2,0),\ \vec{a}_2=(1,\sqrt{3})$. 
Different from the trimerized triangular lattice, due to the existence of flat band, the energy bands are obtained analytically, which reads:
\begin{equation}
\begin{aligned}
    \varepsilon_1&=-(t+t^\prime)\\
    \varepsilon_{2,3}&=\frac{t+t^\prime}{2}\pm\frac{1}{2}\sqrt{9t^2+9t^{\prime 2}+2\left(4K-3\right)tt^\prime}\\
\end{aligned}
\end{equation}
where $K=\cos{[\vec{k}\cdot(\vec{a}_1-\vec{a}_2)]}+\cos{\vec{k}\cdot\vec{a}_2}+\cos{\vec{k}\cdot\vec{a}_1}$,
and they are depicted in Fig.~\ref{Fig:Kagome spectrum}.
The flat band structure remains dispersionless regardless of $t^\prime$, subject to a global energy shift.

Secondly, similar 2nd-order perturbation calculation is applied to  trimerized Kagome lattice case.
The intermediate states calculations are:
\begin{equation}
\begin{aligned}
    H^\prime\ket{S=2}=\frac{\sqrt{2}t^\prime}{3}&\left[\frac{c_{3\uparrow}^\dagger-c_{2\uparrow}^\dagger}{\sqrt{2}}c_{1^\prime\uparrow}^\dagger c_{2^\prime\uparrow}^\dagger c_{3^\prime\uparrow}^\dagger+c_{1\uparrow}^\dagger c_{2\uparrow}^\dagger c_{3\uparrow}^\dagger\frac{c_{1^\prime\uparrow}^\dagger-c_{2^\prime\uparrow}^\dagger}{\sqrt{2}}\right]\ket{\Omega}\\
    H^\prime\ket{S=1}=\frac{t^\prime}{3\sqrt{3}}&\left[\frac{c_{3\uparrow}^\dagger-c_{2\uparrow}^\dagger}{\sqrt{2}}(-c_{1^\prime\uparrow}^\dagger c_{2^\prime\uparrow}^\dagger c_{3^\prime\downarrow}^\dagger+c_{1^\prime\uparrow}^\dagger c_{2^\prime\downarrow}^\dagger c_{3^\prime\uparrow}^\dagger+c_{1^\prime\downarrow}^\dagger c_{2^\prime\uparrow}^\dagger c_{3^\prime\uparrow}^\dagger-2c_{3^\prime\uparrow}^\dagger c_{3^\prime\downarrow}^\dagger c_{1^\prime\uparrow}^\dagger-2c_{3^\prime\uparrow}^\dagger c_{3^\prime\downarrow}^\dagger c_{2^\prime\uparrow}^\dagger)\right.\\
    &\left.-\frac{c_{1\downarrow}^\dagger-c_{2\downarrow}^\dagger}{\sqrt{2}}c_{1^\prime\uparrow}^\dagger c_{2^\prime\uparrow}^\dagger c_{3^\prime\uparrow}^\dagger+c_{1\uparrow}^\dagger c_{2\uparrow}^\dagger c_{3\uparrow}^\dagger\frac{c_{1^\prime\downarrow}^\dagger-c_{2^\prime\downarrow}^\dagger}{\sqrt{2}}\right.\\
    &\left.+(c_{1\downarrow}^\dagger c_{2\uparrow}^\dagger c_{3\uparrow}^\dagger-c_{1\uparrow}^\dagger c_{2\uparrow}^\dagger c_{3\downarrow}^\dagger-c_{1\uparrow}^\dagger c_{2\downarrow}^\dagger c_{3\uparrow}^\dagger-2c_{1\uparrow}^\dagger c_{1\downarrow}^\dagger c_{2\uparrow}^\dagger+2c_{1\uparrow}^\dagger c_{1\downarrow}^\dagger c_{3\uparrow}^\dagger)\frac{c_{1^\prime\uparrow}^\dagger-c_{2^\prime\uparrow}^\dagger}{\sqrt{2}}\right]\ket{\Omega}\\
    H^\prime\ket{S=0}=\frac{t^\prime}{3\sqrt{6}}&\left[\frac{c_{3\uparrow}^\dagger-c_{2\uparrow}^\dagger}{\sqrt{2}}(2c_{1^\prime\downarrow}^\dagger c_{2^\prime\downarrow}^\dagger c_{3^\prime\uparrow}^\dagger-c_{1^\prime\uparrow}^\dagger c_{2^\prime\downarrow}^\dagger c_{3^\prime\downarrow}^\dagger-c_{1^\prime\downarrow}^\dagger c_{2^\prime\uparrow}^\dagger c_{3^\prime\downarrow}^\dagger)\right.\\
    &\left.+\frac{c_{3\downarrow}^\dagger-c_{2\downarrow}^\dagger}{\sqrt{2}}(2c_{1^\prime\uparrow}^\dagger c_{2^\prime\uparrow}^\dagger c_{3^\prime\downarrow}^\dagger-c_{1^\prime\uparrow}^\dagger c_{2^\prime\downarrow}^\dagger c_{3^\prime\uparrow}^\dagger-c_{1^\prime\downarrow}^\dagger c_{2^\prime\uparrow}^\dagger c_{3^\prime\uparrow}^\dagger)\right.\\
    &\left.+(2c_{1\downarrow}^\dagger c_{2\uparrow}^\dagger c_{3\uparrow}^\dagger-c_{1\uparrow}^\dagger c_{2\uparrow}^\dagger c_{3\downarrow}^\dagger-c_{1\uparrow}^\dagger c_{2\downarrow}^\dagger c_{3\uparrow}^\dagger)\frac{c_{1^\prime\downarrow}^\dagger-c_{2^\prime\downarrow}^\dagger}{\sqrt{2}}\right.\\
    &\left.+(2c_{1\uparrow}^\dagger c_{2\downarrow}^\dagger c_{3\downarrow}^\dagger-c_{1\downarrow}^\dagger c_{2\uparrow}^\dagger c_{3\downarrow}^\dagger-c_{1\downarrow}^\dagger c_{2\downarrow}^\dagger c_{3\uparrow}^\dagger)\frac{c_{1^\prime\uparrow}^\dagger-c_{2^\prime\uparrow}^\dagger}{\sqrt{2}}\right]\ket{\Omega}\\
\end{aligned}
\end{equation}
According to the results in SM.~\ref{App:solution of single trimer}, the direct product state of $\frac{c_{1\sigma}^\dagger-c_{2\sigma}^\dagger}{\sqrt{2}}\ket{\Omega}$ and $c_{1\sigma_{1}}^\dagger c_{2\sigma_{2}}^\dagger c_{3\sigma_{3}}^\dagger\ket{\Omega}$ form the intermediate excited state, with total energy $-t$, and the unperturbed energy of the two trimers is $-4t$.
By applying 2nd-order perturbation theory~\ref{eq:2nd_perturbation}, the energy gain of different channels are obtained below:
\begin{equation}
\begin{aligned}
    \Delta E_q^{(2)}&=-\frac{2t^{\prime 2}}{27t}\\
    \Delta E_t^{(2)}&=-\left(\frac{2t^{\prime 2}}{27t}+\frac{88t^{\prime 2}}{81U}\right)\\
    \Delta E_s^{(2)}&=-\left(\frac{2t^{\prime 2}}{27t}+\frac{44t^{\prime 2}}{27U}\right)\\
\end{aligned}
\end{equation}
The effective coupling is
\begin{equation}
    J=\frac{44t^{\prime 2}}{81U}
\end{equation}
which remains AFM for all  $U$ and $t$. 
The physical picture for the difference between the trimerized Kagome lattice and the trimerized triangular lattice is given in the main text.

\section{Preliminary study of metal-insulator transition }\label{greenfunction}
We have developed an analytical theory for the critical interaction $U_c$ in the AFM-FM insulator. 
To support our local spin-moment picture of the AFM-FM transition, the single-electron gaps for a finite-size system have been calculated [Fig.~3(c) in the main text], numerically demonstrating the insulating nature of the AFM phase at intermediate interaction strengths.
Furthermore, a metal-insulator transition is identified in the trimerized model at smaller interaction strengths, for which we aim to provide a preliminary estimate of the phase boundary at first. 
However, no straightforward analytical method exists to determine this transition: 
In the FM regime, the calculated gap for our finite simulated system closely aligns with the thermodynamic gap obtained from band structure calculations, indicating that finite-size effects are negligible for extracting the gap across the $t^{\prime}$-$U$ parameter space. 
Unfortunately, such a reference is unavailable for intermediate interactions within the AFM region. 

Since the band structure of the free terms has no nesting structure, we expect Hubbard's approximation~\cite{hubbard1963electron} for Mott insulator formation can serve as an estimate for the transition.
Below, we give an estimation about where Mott transition takes place.
For simplicity, the following content gives only key steps.

Defining Green's function for operator $c_ic_j^\dagger$:
\begin{equation}
\begin{aligned}
    G(c_{i\sigma},c_{j\sigma}^\dagger;t-t^\prime)&=-i\theta(t-t^\prime)\braket{[c_{i\sigma}(t), c_{j\sigma}^\dagger(t^\prime)]_+}\\
    &:=G_{ij}^\sigma(t-t^\prime)\\
\end{aligned}
\end{equation}
where operators with time parameter are defined in the Heisenberg picture, e.g. $c_{i\sigma}(t):=e^{iHt}c_{i\sigma}e^{-iHt}$, and $\theta(x)$ is a step function.
Applying Fourier transformation to the above Green's function yields
\begin{equation}
    G_{ij}^\sigma(\omega+i\eta)=\frac{i}{2\pi}\int_{-\infty}^{\infty}d(t-t^\prime ) G_{ij}^\sigma(t-t^\prime)e^{i(\omega+i\eta) t}, \quad \eta\rightarrow0^+
\end{equation}
the resulting motion equation of which is
\begin{equation}
    \omega G_{ij}^\sigma(\omega)=\braket{[c_i,c_j^\dagger]_+}+G([c_i,H],c_j^\dagger ;\omega).
    \label{eq:motion equation of Green's function}
\end{equation}
The solution of Eq.~\ref{eq:motion equation of Green's function} relies on the form of higher order Green's function $G([c_i,H],c_j^\dagger ;\omega)$, the motion equation of which satisfies the same form, while $c_{i\sigma}$ is replaced by $[c_{i\sigma},H]$.
Eq.~\ref{eq:motion equation of Green's function} is actually the general formula of the infinite chain equation.
To solve this motion equation, we need to cut off at some order of the motion equation.
Here, to obtain a rough approximation about the metallic-insulating phase transition, we give exact form of motion equations of the first two order:
\begin{equation}
\begin{aligned}
    \omega G_{ij}^\sigma(\omega)&=\delta_{ij}+\sum_j t_{ij}G_{ij}^\sigma(\omega) + UG(n_{i\bar{\sigma}}c_{i\sigma},c_{i\sigma}^\dagger;\omega)\\
    =&\delta_{ij}+\sum_j t_{ij}G_{ij}^\sigma(\omega) +U\sum_{\alpha=\pm} n^\alpha G(n_{i\bar{\sigma}}c_{i\sigma},c_{i\sigma}^\dagger;\omega)\\
    =&\delta_{ij}+\sum_j t_{ij}G_{ij}^\sigma(\omega) +U\sum_{\alpha=\pm} F_{ij}^{\sigma\alpha}(\omega)\\
    \omega F_{ij}^{\sigma\alpha}(\omega)=&\braket{n_{i\bar{\sigma}}^\alpha}\left[\delta_{ij}+\sum_l t_{il}G_{ij}(\omega)\right]+\varepsilon_\alpha F^{\sigma\alpha}_{ij}(\omega)+\sum_l t_{il}G\left((n^\alpha_{i\bar{\sigma}}-\braket{n^\alpha_{\bar{\sigma}}})c_{l\sigma},c_{j\sigma}^\dagger;\omega\right)\\
    &+\alpha\sum_l t_{il}\left[G(c_{i\bar{\sigma}}^\dagger c_{l\bar{\sigma}}c_{i\sigma},c_{j\sigma}^\dagger;\omega)-G(c_{l\bar{\sigma}}^\dagger c_{i\bar{\sigma}}c_{i\sigma},c_{j\sigma}^\dagger;\omega)\right]\\
\end{aligned}
\end{equation}
where for simplicity we define $n^+_{i\sigma}=n_{i\sigma},n^-_{i\sigma}=1-n_{i\sigma},\varepsilon_+=\frac{1}{2}U,\varepsilon_-=-\frac{1}{2}U$ and the Hamiltonian is simplified as $H=\left(t\sum_{\braket{ij},\sigma}c_{i\sigma}^\dagger c_{j\sigma} + t^\prime\sum_{\braket{\braket{i^\prime j^\prime}},\sigma}c_{i^\prime\sigma}^\dagger c_{j^\prime\sigma}+h.c.\right)+U\sum_in_{i\uparrow}n_{i\downarrow}\stackrel{\Delta}=\sum_{ij}t_{ij}c_i^\dagger c_j+\frac{U}{2}\sum_{i\sigma}n_{i\sigma}n_{i\bar{\sigma}}$.
Then by ignoring the last two terms separately and by truncating the hierarchy to obtain a closed set of equations, we could obtain an approximate solution of $G_{ij}^{\sigma}(\omega)$, the forms of which are the same
\begin{equation}
\begin{aligned}
    G_{k}^\sigma(\omega)&=\frac{1}{F^\sigma_a(\omega)-\varepsilon_k}\\
    \frac{1}{F_a^\sigma(\omega)}&=\frac{\omega-\left(\braket{n^+_{\sigma }}\varepsilon_-+\braket{n^-_{\sigma }}\varepsilon_+\right)-\Omega_{\sigma a}}{\left[\omega-\varepsilon_--\braket{n^+_{\sigma }}\Omega_{\sigma a}\right]\left[\omega-\varepsilon_+-\braket{n^-_{\sigma }}\Omega_{\sigma a}\right]-\braket{n_{\sigma }^-}\braket{n^+_{\sigma }}\Omega_{\sigma a}^2}\\
\end{aligned}
\label{eq:solution of Green's function}
\end{equation}
where $\varepsilon_k$ is the structure of the free energy band calculated by the Fourier transformation of $\sum_{ij,\sigma}t_{ij}c_{i\sigma}^\dagger c_{j\sigma}$, and $\Omega_{\sigma a}$ is defined as follows:
\begin{equation}
\begin{aligned}
    G_{ii}^\sigma(\omega)&=\frac{1}{F_1^\sigma(\omega)-\Omega_{\sigma 1}(\omega)}\\
    \Omega_{\sigma 2}(\omega)&=\Omega_{\bar{\sigma}1}(\omega)-\Omega_{\bar{\sigma}1}(-\omega)
\end{aligned}
\end{equation}
Combining the two solutions, we obtain the final solution, which is the same as Eq.~\ref{eq:solution of Green's function} with $\Omega_{\sigma}(\omega)=\Omega_{\sigma 1}(\omega)+\Omega_{\sigma2}(\omega)$.
The self-consistent equation comes from
\begin{equation}
\begin{aligned}
    \frac{1}{F^\sigma(\omega)-\Omega_{\sigma}(\omega)}=G_{ii}^\sigma(\omega)&=\frac{1}{\mathcal{N}}\sum_k G_k^\sigma(\omega)\\
    &=\int \frac{p(\varepsilon)d\varepsilon}{F^\sigma(\omega)-\varepsilon}
\end{aligned}
\end{equation}
where $p(\varepsilon)$ is DOS of free energy band.
The DOS is approximated to be $p(\varepsilon)=\frac{4}{\pi W}\sqrt{1-\left(\varepsilon/\frac{1}{2}W\right)^2}$ only if $|\varepsilon|\le\frac{1}{2}W$, where $W$ is the bandwidth, else $p(\varepsilon)=0$.
Here, the bandwidth $W$ is defined as the combined width of two lower overlapping bands.
It should be consistent with Eq.~\ref{eq:solution of Green's function}.
Finally, the critical interaction strength at which the bandgap vanishes is:
\begin{equation}
    \frac{U_c}{W}\approx\frac{\sqrt{3}}{2}
\end{equation}
from the self-consistent equation. 
Finally, by computing $W$ numerically ($W\sim 4t^\prime/t$ for $t^\prime/t \le 0.6$), the qualitative transition phase boundary of the metal-insulator is sketched, as depicted in Fig. 3 (a) in the main text.

\section{DMRG convergence analysis}
\label{DMRG_convergence}

\begin{figure}[htp]
    \centering
    \includegraphics[width=0.8\linewidth]{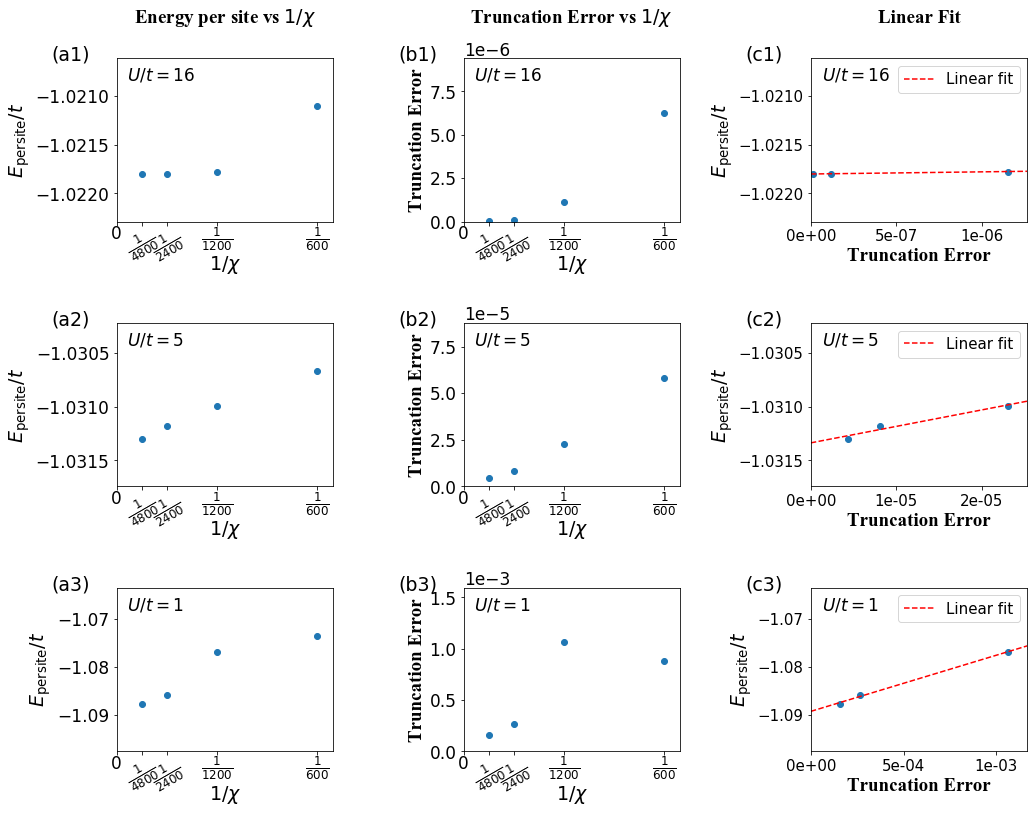}
    \caption{Scaling of the DMRG simulation results for the system defined on a tilted cylinder with the size $L_x=\infty$ and $L_y=4$ and each unit cell as a trimer. 
    The parameter values are $t^\prime/t=0.2$; $U/t=16, 5$ and $1$ for (a1) to (c1), (a2) to (c2), and (a3) to (c3), respectively, which represent the FM insulating, AFM insulating, and non-magnetic regions, respectively. 
    (a1-a3) The ground state energy per site ($E_{\text{per site}}/t$) and (b1-b3) the truncation error with respect to the inverse of bond dimensions $1/\chi$. 
    (c1-c3) The linear fit between the ground state energy and the truncation error, where three data points ($\chi=1200,2400,4800$) are employed to extrapolate $E_{\text{per site}}/t$ in the limit of $\chi\rightarrow\infty$. }
    \label{fig:DMRG_convergence}
\end{figure}

The DMRG convergence results are presented in Fig.~\ref{fig:DMRG_convergence}, based on the analysis of the ground-state energies $E_{\text{per site}}$ and truncation errors with respect to the inverse of bond dimension $1/\chi$. 
We show the data analysis in Fig.~\ref{fig:DMRG_convergence}, for a $\infty \times 4 \times 3$ cylinder which is the largest geometry we consider.

The analysis is presented in three regions: the ferromagnetic insulating region (FM) ($U/t=16$), the antiferromagnetic insulating region (AFM) ($U/t=5$), and the non-magnetic region ($U/t=1$), all at a fixed $t^\prime/t=0.2$. 

The precision of data is characterized from three perspectives: 
\begin{enumerate}
    \item How the simulated energy ($E_{\mathrm{per\ site}}/t$) approaches the precise value by decreasing the dimension of the inverse bond ($1/\chi$). 
    The trend is shown in Fig.~\ref{fig:DMRG_convergence} (a1)-(a3). 
    \item SVD truncation errors of the two-site DMRG algorithm are used to characterize convergence, shown in Fig.~\ref{fig:DMRG_convergence} (b1)-(b3). 
    \item A linear extrapolation is applied between $E_{\mathrm{per\ site}}/t$ and the truncation error~\cite{PhysRevLett.99.127004}, which also provides the extrapolation of the ground state energy in the $1/\chi\to 0$ limit. The extrapolation is plotted in Fig.~\ref{fig:DMRG_convergence} (c1)-(c3). 
\end{enumerate}

For the data in the insulator region of interest, our DMRG results show high precision in convergence. 
Specifically, in the FM insulating region, the truncation error has been reduced below $10^{-8}$, and the difference between the best DMRG simulated energy and the extrapolated energy is negligible, at the margin of approximately $10^{-7}$. 
For data in the AFM insulating region, the truncation error is reduced to the $10^{-6}$ level, with the difference between the computed and extrapolated energies at $\sim 10^{-5}$. 
As expected, convergence is comparatively slower in the weak interaction regime ($U/t=1$), leading to larger truncation errors $\sim 10^{-4}$. 
This makes the studies of metallic phases and the metal-insulator transition in a weak interaction challenging. 
Nevertheless, our data for $U/t\ge 2$ show a fast convergence trend and high precision in energy. 
The precision of our data for finite geometry is higher than the results analyzed above. 
The precision is also sufficient for correlation function data of interest; see below.


\section{DMRG data for spin-spin correlation functions}

\begin{figure}[htp]
    \centering
    \includegraphics[width=0.7\linewidth]{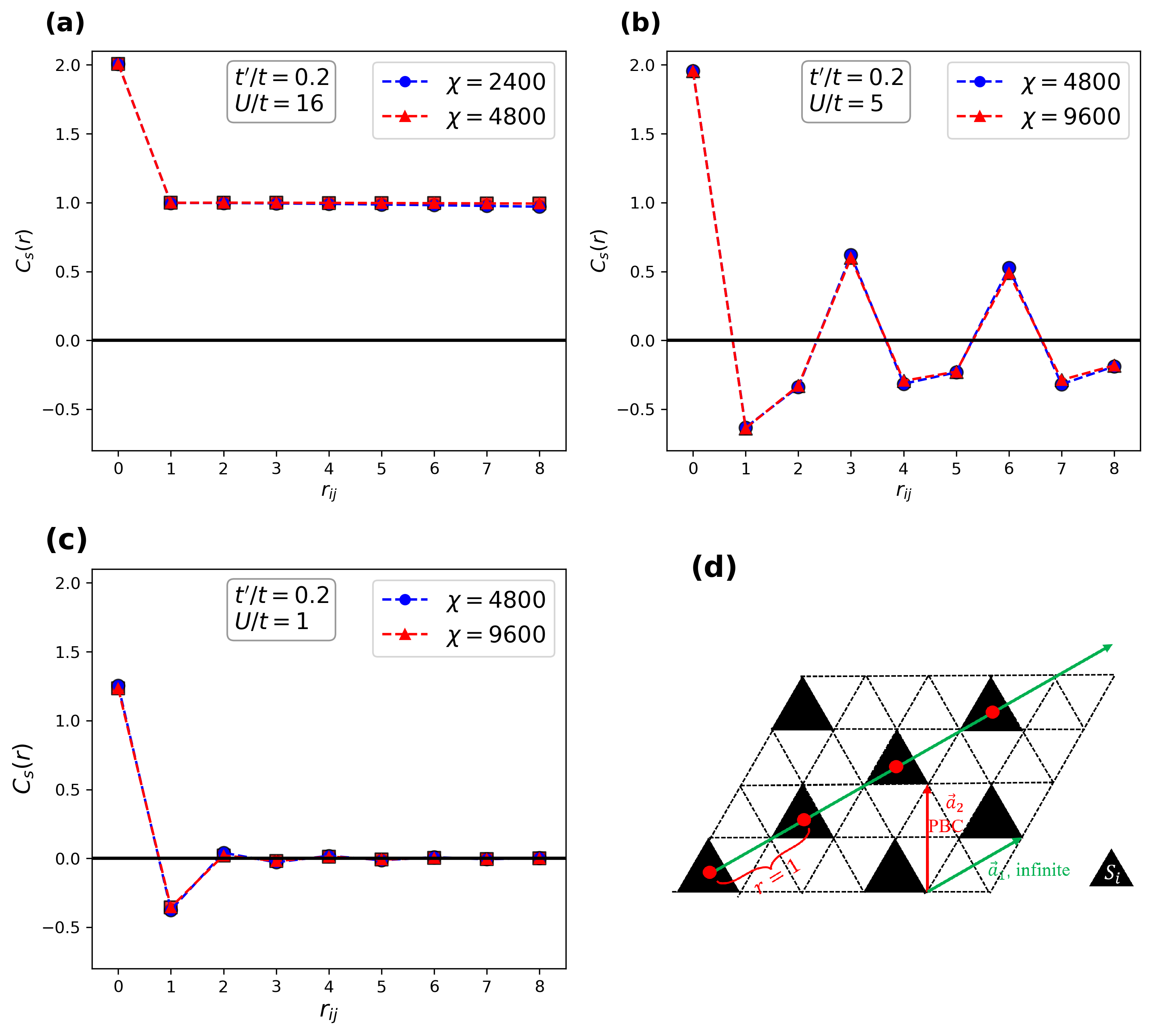}
    \caption{Results for trimer spin-spin correlation functions for (a) $U/t=16$, (b) 5 and (c) 1, respectively. 
    Two different bond dimensions are used in each plot with $\chi=4800,2400$ in (a) and $\chi=9600,4800$ in (b) and (c). Data for different bond dimensions shows convergence in precision. 
    The dashed lines are guides to eyes. 
    (d) Each trimer is indicated as a black triangle. 
    The blue arrow indicates the axial direction along which the spin correlation functions are calculated.
    }
    \label{fig:corr_fun_spin}
\end{figure}

In this section, we present the trimer spin correlation functions $C_s(r_{ij})$, at $t^\prime/t=0.2, U/t=16,5,1$, as shown in Fig.~\ref{fig:corr_fun_spin} (a)-(c). 
In this and the following sections, we analyze the DMRG data for $\infty \times 3 \times 3$ cylinder, meaning that there are three trimers around the cylinder. 

The trimer spin correlation functions are defined as:
\begin{equation}
    C_s(r_{ij})=\braket{\vec{S}_i\cdot \vec{S}_j}
    \label{eq:spin correlation function}
\end{equation}
where $\vec{S}_i$ and $\vec{S}_j$ are the total spin operators of the trimers $i$ and $j$, respectively, and $r_{ij}$ is defined as the distance between two equivalent positions of two trimers $i$ and $j$, along the axial direction shown in Fig.~\ref{fig:corr_fun_spin} (b).

The inferences of AFM order and of FM order based on the DMRG data are quite different. 
As shown in the main text, the FM order parameter can be directly estimated by calculating the total spin. 
Due to the fact that net magnetic moment enters the correlation function as the disconnected part at $T=0$, the correlation function in FM region does not decay to zero even at the large distance limit. 
In contrast, the AFM order parameter is not a conserved charge, unlike the one in the FM sector. 
Thus, with a finite size or finite width, there is no spontaneous symmetry breaking to directly reveal the order parameter. 
Specifically, for finite-width cylinder geometry, the correlation functions decay to zero at the large distance limit along the axial direction. 
Nevertheless, the correlation function can still reveal the spin density wave pattern of the AFM order, and the correlation function is also expected to decay relatively slow and maintain large value at least up to the cylinder width scale to reflect the spin density waves.

According to both Fig. 3 (a) and (b) in the main text, the state of $t^\prime/t=0.2,U/t=16$ lies in the fully spin polarized region. 
The corresponding spin correlation function $C_s (r_{ij})$ is shown in Fig.~\ref{fig:corr_fun_spin} (a). 
For $r=0$, $C_s (0)=\braket{S_i^2}=2$ since the system is FM insulating and each trimer forms a spin triplet, e.g. $S=1$. 
$C_s (r_{ij})=1$ exhibits the plateau for $r_{ij}\ge1$, which is consistent with the fully polarized FM insulating state. 

For $U/t=5$, the spin correlation function calculated by DMRG shows oscillations with an approximate period of 3, which is consistent with the $120^\circ$ AFM ordering in triangular lattice. 
The decay is slow, which implies that in the 2D thermodynamic limit, the system would develop AFM long-range ordering or possess strong AFM quantum fluctuations. 

As for $U/t=1$, the DMRG spin correlation function decays much faster than $U/t=5$. 
This indicates the absence of magnetic ordering in this weak-coupling region.

The above results support our prediction that antiferromagnetic and ferromagnetic exchange dominate at the intermediate and large $U$, respectively.



\section{DMRG data for single-electron correlation length in the metallic region}

\begin{figure}[htp]
    \centering
    \includegraphics[width=0.7\linewidth]{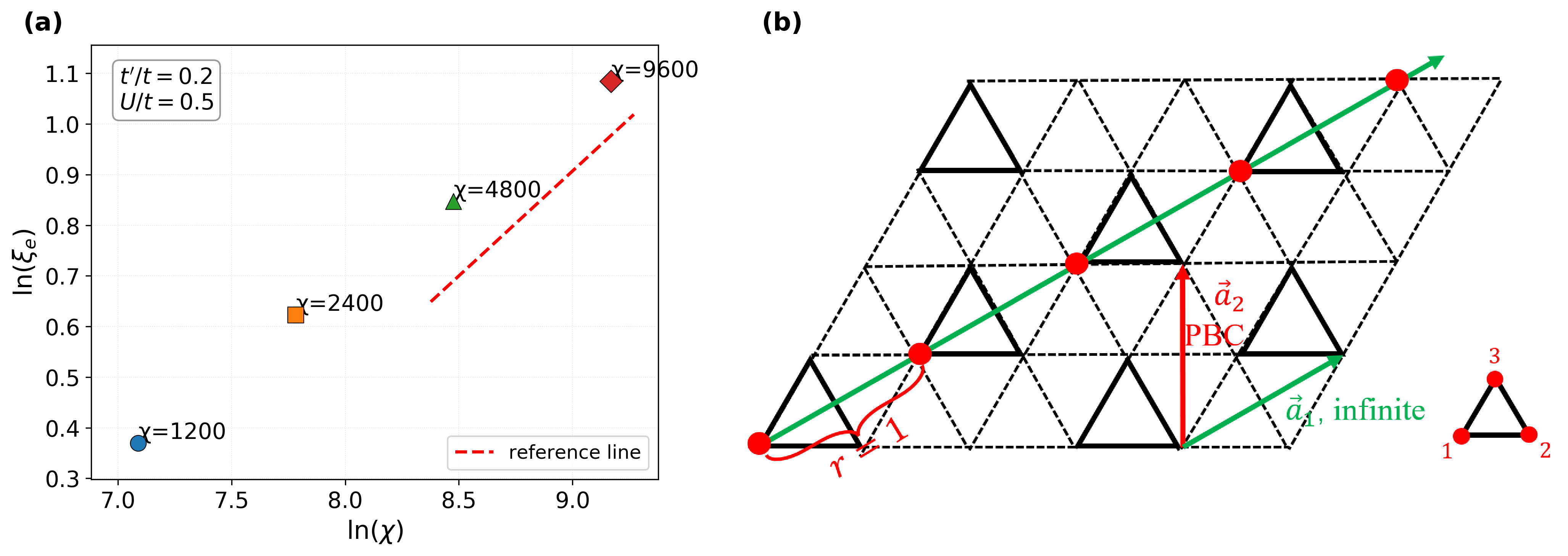}
    \caption{(a) Finite bond-dimension correlation length scaling  for $\xi_e$ at $t^\prime/t=0.2, U/t=0.5$, with bond dimension $\chi=1200,2400,4800,9600$.  The dashed line is a reference slope for  correlation length scaling  near the scaling limit $\chi \rightarrow \infty$. Under the proposition of Ref.~\onlinecite{pollmann2009}, the reference slope is expected if three transverse momenta cut through the Fermi surface as the $U=0$ case for $\infty \times 3 \times 3$ cylinder. 
    The transverse momenta $0, \frac{2\pi}{3},\frac{4\pi}{3}$ are quantized due to the geometry.
    (b) The tilted cylinder geometry with the green axis indicating the direction along which electron correlations are calculated.}
    \label{fig:cor_len_ele}
\end{figure}

In this section, we present the electron correlation length as a function of the bond dimension $\chi$ at $t^\prime/t=0.2,U/t=0.5$, as shown in Fig.~\ref{fig:cor_len_ele} (a). 
The electron correlation length $\xi_e$ is defined from the single-electron equal-time correlation function $C_e (i,j)$ along the axial direction of the cylinder, as shown in Fig.~\ref{fig:cor_len_ele} (b). 
$C_e (i,j)$ is defined as,
\begin{equation} 
    C_e(i,j) = \frac{1}{2}\sum_{\sigma}\langle c_{ia \sigma} c_{ja \sigma}^{\dagger} \rangle \propto e^{-|\mathbf{r}_{ia}-\mathbf{r}_{ja}| / \xi_e}, 
\end{equation}\label{ce}
where $i,j$ label the location of the trimer and $a=1,2,3$ labels three different sites within the trimer. 
The choice of sites and the direction of the correlation function is plotted in Fig.~\ref{fig:cor_len_ele} (b). 
The length unit is chosen as the lattice constant. 
In addition to fitting, the correlation length can be read from the dominant values of the transfer matrix spectrum, because the correlation function of an iMPS is numerically computed from the transfer matrix as the sum of exponentially decaying components~\cite{zauner2015transfer,rams2018}.

For a gapless system, the DMRG results exhibit the so-called finite-correlation scaling or finite-entanglement scaling~\cite{rams2018, pollmann2009}, characterized by $\xi_e(\chi)\propto\chi^\kappa$ in the large $\chi$ limit, where $\kappa$ is a constant. 
Therefore, for a quasi-one-dimensional metallic state, DMRG estimated electron correlation length $\xi_e(\chi)$ is expected to increase with increasing $\chi$.

Fig~\ref{fig:cor_len_ele} (a) shows such behavior of the $\xi_e(\chi)$ for a state with weak interaction ($t'/t=0.2, U=0.5$); a reference dashed line is plotted. Under the proposition of Ref.~\onlinecite{pollmann2009}, the reference slope is expected for $\chi \rightarrow \infty$ if three transverse momentum cut through Fermi surface as the $U=0$ case for $\infty \times 3 \times 3$ cylinder. 
The transverse momentum $0, \frac{2\pi}{3},\frac{4\pi}{3}$ is quantized due to the geometry. 
This indicates that the system is in the gapless metallic state at $t'/t=0.2$ and $U/t=0.5$.

\section{Correlation length data in the AFM insulator region}

In the preceding two sections, we have discussed the electron-electron correlation length in the metallic region and trimer spin-spin correlation functions for the system on a cylinder geometry. 
For $U=16$ and $U=5$, these data have errors smaller than marker size within the plotted region. 
Beyond the plotted region, the long-range limit is governed by cylindrical quasi-one-dimensional physics rather than our concerned two-dimensional physics. 

\begin{figure}[htp]
    \centering
    \includegraphics[width=0.8\linewidth]{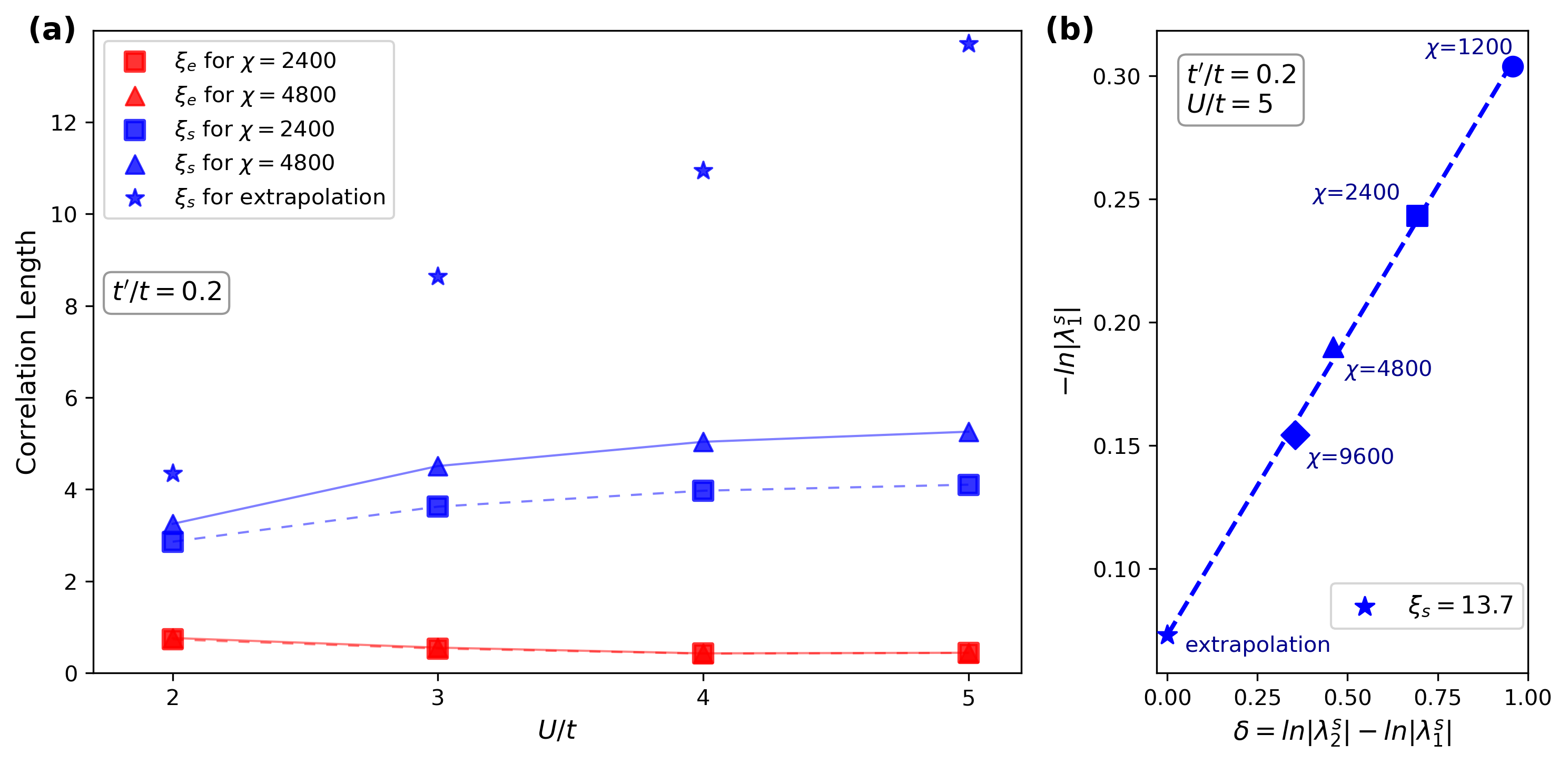}
    \caption{(a) Electron correlation length (red) and spin correlation length (blue) versus $U/t$ at $t^\prime/t=0.2$. 
    The system size is $L_x\times L_y\times 3$ with $L_x=\infty$ and $L_y=3$, and a trimer as a unit cell. 
    Square and triangle data points represent bond dimension of $\chi=2400$ and $4800$, respectively. 
    (b) Extrapolation for $\xi_s$, with the relation $\xi_s=-1/\ln{|\lambda_1^s|}$. The parameter value are $t^\prime/t=0.2, U/t=5$ and bond dimensions $\chi=1200,2400,4800,9600$. 
    The method uses the first two dominant eigenvalues of transfer matrix $|\lambda_1^s|, |\lambda_2^s|$.
    }
    \label{fig:corr_len}
\end{figure}

In this section [Fig.~\ref{fig:corr_len} (a)], we compare the spin correlation length $\xi_s$ and the electron correlation length $\xi_e$ for $t^\prime/t=0.2, U/t=2\sim 5$ in the AFM region.

The electron correlation lengths are marked by red symbols, and symbols with doubling the bond dimension overlap, indicating indicating good convergence.
At $U/t>2$, the $\xi_e$ values are very small below 1 (please note that we set the inter-trimer distance as $1$ which is $\sqrt{3}$ times larger than the edge length of a trimer.)
This feature means that electrons are very localized at the scale of one trimer, which is a signature of insulating state.

The spin correlation lengths $\xi_s$ are marked by blue symbols. 
They are extracted from the eigenvalue spectra of the transfer matrix. 
The eigenvalues are labeled by spin and charge quantum numbers, which allow us to extract $\xi_s$ from the dominant eigenvalues in the relevant sectors via $\xi_s=-\ln|\lambda^s_1|$~\cite{zauner2015transfer}. 
(This works for $\xi_e$ as well.) 
As the bond dimension is doubled, $\xi_s$ increases. 
The data at $\chi=4800$ represent a lower bound for $\xi_s$, hence $\xi_s$ is much larger than $\xi_e$. 

In Fig.~\ref{fig:corr_len} (b), the extrapolation for spin correlation length is performed~\cite{rams2018} for $U/t=5$, which gives a large value around $13.7$.
Due to the quasi-1D geometry, the ground state should be disordered due to quantum AFM fluctuations. 
For spin-one quasi-one-dimensional systems, small spin gaps may open.
As for the 2D thermodynamic limit, we expect $\xi_s$ to be divergent as $\chi\rightarrow\infty$, indicating the gapless nature in the spin channel. 
Hence, it is reasonable to expect that the ground state should be AFM long-range ordered with the $120^\circ$ pattern of the triangular lattice. 

We do not attempt to determine the metal-insulator transition lines using DMRG. 
As we see in Fig.~\ref{fig:DMRG_convergence} and Fig.~\ref{fig:cor_len_ele},  DMRG requires fast-growing bond dimensions to improve the accuracy for metallic or small-charge-gap insulating states. 
The determination of the metal-insulator transition has a further serious finite-size effect issue. 
Instead of determining the transition, we denote data points for states with extremely short electron-electron correlation lengths in the phase diagram of the main text.

\bibliographystyle{prsty}
\bibliography{ref}